\documentclass{article}
\usepackage{epsfig}
\usepackage{indentfirst}
\usepackage[psamsfonts]{amssymb}
\font\smallroman=cmr10 at 8pt

\newtheorem{theorem}{Theorem}
\newtheorem{definition}{Definition}

\newtheorem{axiom}{Axiom}

\date{}
\begin{document}
\title {Quantum Mechanics: Structures, \\ Axioms and Paradoxes\footnote{Published as: Aerts,
D., 1999, ``Quantum Mechanics; Structures, Axioms and Paradoxes", in {\it Quantum Structures
and the Nature of Reality: the Indigo book of the Einstein meets Magritte series}, eds. Aerts,
D. and Pykacz, J., Kluwer Academic, Dordrecht.}}
\author{Diederik Aerts}
\maketitle


\centerline{Center Leo Apostel,}
\centerline{Brussels Free University,}
\centerline{Krijgskundestraat 33,}
\centerline{1160 Brussels, Belgium.}
\centerline{e-mail: diraerts@vub.ac.be}

\begin{abstract}
\noindent We present an analysis of quantum
mechanics and its problems and paradoxes
taking into account the results that have been
obtained during the last two decades by investigations in the field of `quantum structures research'. We concentrate mostly on the results
of our group FUND at Brussels Free University. By means of a spin ${1 \over 2}$ model where the quantum probability is generated by the
presence of fluctuations on the interactions between measuring apparatus and physical system, we show that the quantum structure can find
its origin in the presence of these fluctuations. This appraoch, that we have called the `hidden measurement approach', makes it possible
to construct systems that are in between quantum and classical. We show that two of the traditional axioms of quantum axiomatics are not
satisfied for these `in between quantum and classical' situations, and how this structural shortcoming of standard quantum mechanics is at
the origin of most of the quantum paradoxes. We show that in this approach the EPR paradox identifies a genuine incompleteness of standard
quantum mechanics, however not an incompleteness that means the lacking of hidden variables, but an incompleteness pointing at the
impossibility for standard quantum mechanics to describe separated quantum systems. We indicate in which way, by redefining the meaning of
density states, standard quantum mechanics can be completed. We put forward in which way the axiomatic approach to quantum mechanics can
be compared to the traditional axiomatic approach to relativity theory, and how this might lead to studying another road to unification of
both theories.

\end{abstract}

\section{Introduction}
\noindent In this article we present an analysis of quantum
mechanics and its problems and paradoxes
taking into account some of the results and insights that have been
obtained during the last two decades by investigations that are
commonly classified in the field of `quantum structures research'. We
will concentrate on these aspects of quantum mechanics that have been
investigated in our group in FUND at the Free University of
Brussels\footnote{The actual members of our research group FUND -
Foundations of the Exact Sciences - are: Diederik Aerts, Bob Coecke,
Thomas Durt, Sven Aerts, Frank Valkenborgh, Bart D'Hooghe, Bart Van
Steirteghem and Isar Stubbe.}. We try to be as clear and
self contained as possible: firstly because the article is also 
aimed at scientists not specialized in quantum mechanics,
and secondly because we believe that some of the results and insights
that we have obtained in Brussels present the deep problems of quantum
mechanics in a simple and new way.

\par The study of the structure of quantum mechanics is almost as old
as quantum mechanics itself. The fact that the two early
versions of quantum mechanics - the matrix mechanics of Werner
Heisenberg and the wave mechanics of Erwin Schr{\"o}dinger -
were shown to be structurally equivalent to what is now
called standard quantum mechanics, made it already clear in the
early days that the study of the structure itself would
be very important.
The foundations of much of this structure are already present
in the book of John Von Neumann [1], and if we
refer to standard quantum mechanics we mean the formulation of the
theory as it was first presented there in a complete way.

\par Standard quantum mechanics makes use of a sophisticated
mathematical apparatus, and this is one of the reasons that it is not
easy to explain it to a non specialist audience. Upon reflecting how
we would resolve this `presentation' problem for this paper we have
chosen the following approach: most, if not all, deep quantum
mechanical problems appear already in full, for the case of the
`most simple' of all quantum models, namely the model for the spin of a
spin${1 \over 2}$ quantum particle. Therefore we have chosen to
present the technical aspects of this paper as much as possible for
the description of this most simple quantum model, and to expose the
problems by making use of its quantum mechanical and quantum structural
description. The advantage is that the structure needed to explain the
spin model is simple and only requires a highschool background in
mathematics.

\par
The study of quantum structures has been motivated mainly by two
types of shortcomings of standard quantum mechanics. (1) There is no
straightforward physical justification for the use of the abstract
mathematical apparatus of quantum mechanics. By introducing an
axiomatic approach the mathematical apparatus of standard quantum
mechanics can be derived from more general structures that can be
based more easily on physical concepts (2) Almost none of the
mathematical concepts used in standard quantum mechanics are
operationally defined. As a consequence there has also been a great
effort to elaborate an operational foundation.

\section{Quantum structures and quantum logic}

\noindent Relativity theory, formulated in
great part by one person, Albert Einstein, is founded on the concept
of `event', which is a concept that is physically well defined and
understood. Within relativity theory itself, the events are
represented by the points of a four dimensional space-time continuum.
In this way, relativity theory has a well defined physical and
mathematical base. 

\par
Quantum mechanics on the contrary was born in
a very obscure way. Matrix mechanics was constructed by Werner
Heisenberg in a mainly technical effort to explain and describe the
energy spectrum of the atoms. Wave mechanics, elaborated by Erwin
Schr{\"o}dinger, seemed to have a more solid physical base: a general
idea of wave-particle duality, in the spirit of Louis de Broglie or
Niels Bohr. But then Paul Adrien Maurice Dirac and later John Von
Neumann proved that the matrix mechanics of Heisenberg and the wave
mechanics of Schr{\"o}dinger are equivalent: they can be
constructed as two mathematical representations of one and the same
vector space, the Hilbert space. This
fundamental result indicated already that the `de Broglie wave' and the
`Bohr wave' are not physical waves and that the state of a quantum
entity is an abstract concept: a vector in an abstract vector space.

\par Referring again to what we mentioned to be the two main reasons
for studying quantum structures, we can state now more clearly: the
study of quantum structures has as primary goal the elaboration of
quantum mechanics with a physical and mathematical base that is as
clear as the one that exists in relativity theory. We remark
that the initial aim of quantum structures research was not to
`change' the theory - although it ultimately proposes a fundamental
change of the standard theory as will be outlined in this paper - but
to elaborate a clear and well defined base for it.

\par For this purpose it is necessary to introduce clear and
physically well defined basic concepts, like the events in the theory
of relativity, and to identify the mathematical structure that these
basic concepts have to form to be able to recover standard quantum
mechanics.

\par In 1936, Garret Birkhoff and John Von Neumann, wrote an article
entitled ``The logic of quantum mechanics". They show that if one
introduces the concept of `operational proposition' and its
representation in standard quantum mechanics by an orthogonal
projection operator of the Hilbert space,
it can be shown that the set of the `experimental propositions' does
not form a Boolean algebra, as it the case for the set of
propositions of classical logic [2]. As
a consequence of this article the field called `quantum logic' came
into existence: an investigation on the logic of quantum mechanics.

\par An interesting idea was brought forward. Relativity theory
is a theory based on the concept of `event' and a mathematical
structure of a four dimensional space-time continuum. This
space-time continuum contains a non Euclidean geometry. Could it
be that the article of Birkhoff and Von Neumann indicates that
quantum mechanics should be based on a non Boolean logic in the same
sense as relativity theory is based on a non Euclidean geometry?
This is a fascinating idea, because if quantum mechanics
were based on a non Boolean logic, this would perhaps
explain why paradoxes are so abundant in quantum mechanics: the
paradoxes would then arise because classical Boolean logic is used to
analyze a situation that intrinsically incorporates a non classical,
non Boolean logic.

\par Following this idea quantum logic was developed as a
new logic and also as
a detailed study of the logico-algebraic structures that are
contained in the mathematical apparatus of quantum mechanics. The
systematic study of the logico-algebraic structures related to quantum
mechanics was very fruitful and we refer to the paper that David
Foulis published in this book for a good historical account [3]. On
the philosophical question of whether quantum logic constitutes a
fundamental new logic for nature a debate started. A good overview of
this discussion can be found in the book by Max Jammer [4].

\par We want to put forward our own personal opinion about this
matter and explain why the word `quantum logic'
was not the best word to choose to indicate the
scientific activity that has been taking place within this field. If
`logic', following the characterization of Boole, is the
formalization of the `process of our reflection',
then quantum logic is not a new logic. Indeed, we obviously reflect
following the same formal rules whether we reflect about classical
parts of reality or whether we reflect about quantum parts of
reality. Birkhoff and Von Neumann, when they wrote their article in
1936, were already aware of this, and that is why they
introduced the concept of `experimental proposition'. It could indeed
be that, even if we reason within the same formal structure about
quantum entities as we do about classical entities, the structure of
the `experimental propositions' that we can use are different in both
cases. With experimental proposition is meant a proposition that is
connected in a well defined way with an experiment that can test this
proposition. We will explicitly see later in this paper that there is
some truth in this idea. Indeed, the set of experimental propositions
connected to a quantum entity has a different structure than the set
of experimental propositions connected to a classical entity. We
believe however that this difference in structure of the sets of
experimental propositions is only a little piece of the problem, and
even not the most important one\footnote{We can easily show for
example that even the set of experimental propositions of a
macroscopic entity does not necessarily have the structure of a
Boolean algebra. This means that the only fact of limiting oneself to
the description of the set of `experimental' propositions already
brings us out of the category of Boolean structures, whether the
studied entities are microscopic or macroscopic [5]}. It is our
opinion that the difference between the logico-algebraic structures
connected to a quantum entity and the logico-algebraic structures
connected to a classical entity is due to the fact that the structures
of our `possibilities of active experimenting' with these entities is
different. Not only the logical aspects of these possibilities of
active experimenting but the profound nature of these possibilities of
active experimenting is different. And this is not a subjective matter
due to, for example, our incapacity of experimenting actively in the
same way with a quantum entity as with a classical entity. It is the
profound difference in nature of the quantum entity that is at the
origin of the fact that the structure of our possibilities of active
experimenting with this entity is different\footnote{We have to remark
here that we do not believe that the set of quantum entities and the
set of classical entities correspond respectively to the set of
microscopic entities and the set of macroscopic entities as is usually
thought. On the contrary, we believe, and this will become clear step
by step in the paper that we present here, that a quantum entity
should best be characterized by the nature of the structure of the
possibilities of experimentation on it. In this sense classical
entities show themselves to be special types of quantum entities,
where this structure, due to the nature of the entity itself, takes a
special form. But, as we will show in the paper, there exists
macroscopic real physical entities with a quantum structure.}. We
could proceed now by trying to explain in great generality what we
mean with this statement and we refer the reader to [6] for such a
presentation. In this paper we will explain what we mean mostly by
means of a simple example.

\section{The example: the quantum machine}

\noindent
As we have stated in the introduction, we will analyze the problems
of quantum mechanics by means of simple models. The first model that
we will introduce has been proposed at earlier occasions (see [5],
[7], [8] and [9]) and we have named it the `quantum machine'. It will
turn out to be a real macroscopic mechanical model for the spin of a
spin${1
\over 2}$ quantum entity. We will only need the mathematics of
highschool level to introduce it. This means that also the
readers that are not acquainted with the sophisticated mathematics of
general quantum mechanics can follow all the calculations, only
needing to refresh perhaps some of the old high school mathematics.

\par We
will introduce the example of the quantum machine model step by step,
and before we do this we need to explain shortly how we represent our
ordinary three dimensional space by means of a real three dimensional
vector space.

\subsection{The mathematical representation of three
dimensional space} \label{space}
\noindent
We can represent the three dimensional Euclidean space, that is also
the space in which we live and in which our classical macroscopic
reality exists, mathematically by means of a three dimensional real
vector space denoted $\Bbb{R}^3$. We do this by
choosing a fixed origin
$0$ of space and representing each point $P$ of space as a vector $v$
with begin-point $0$ and end-point $P$ (see Fig 1). Such a vector $v$
has a direction, indicated by the arrow, and a length, which is the
length of the distance from $0$ to the point $P$. We denote the
length of the vector $v$ by $|v|$.
\vskip 0.5 cm

\hskip 3.5 cm \includegraphics{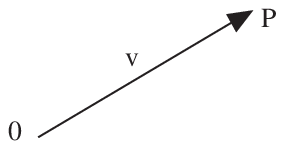}

\begin{quotation}
\noindent \baselineskip= 7pt \smallroman Fig 1 : A mathematical representation
of the three dimensional Euclidean space by means of a three
dimensional real vector space. We
choose a fixed point
$\scriptstyle 0$ which is the origin. For an arbitrary point
$\scriptstyle P$ we define the vector $\scriptstyle v$ that
represents the point.
\end{quotation}

\medskip
\noindent {\it i) The sum of vectors}
\smallskip

\noindent
We introduce operations that can be performed with these vectors that
indicate points of our space. For example the sum of two
vectors $v, w \in \Bbb{R}^3$ representing two points $P$ and
$Q$ is defined by means of the parallelogram rule (see Fig 2). It is
denoted by $v+w$ and is again a vector of $\Bbb{R}^3$.

\vskip 0.5 cm

\hskip 3.5 cm \includegraphics{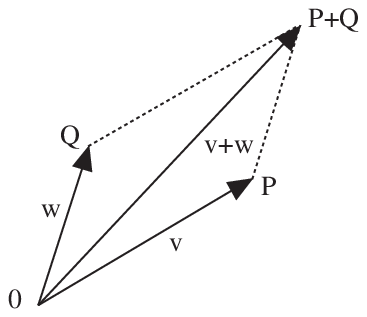}

\begin{quotation}
\noindent \baselineskip= 7pt \smallroman Fig 2 : A representation
of the sum of two vectors $\scriptstyle v$ and $\scriptstyle w$,
denoted by $\scriptstyle v+w$.
\end{quotation}

\medskip
\noindent {\it ii) Multiplication of a vector by a real number}
\smallskip

\noindent
We can also define the multiplication of a vector $v \in \Bbb{R}^3$ by a 
real number $r \in \Bbb{R}$, denoted by $rv$. It is
again a vector of $\Bbb{R}^3$ with the same direction and the
same origin $0$ and with length given by the original length of the
vector multiplied by the real number $r$ (see Fig 3).

\vskip 0.5 cm

\hskip 3.5 cm \includegraphics{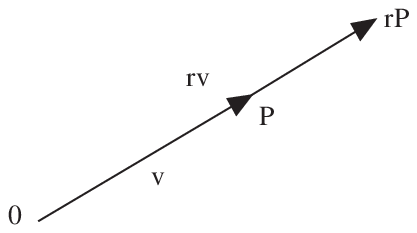}

\begin{quotation}
\noindent \baselineskip= 7pt \smallroman Fig 3 : A representation
of the product of a vector $\scriptstyle v$ with a real number
$\scriptstyle r$, denoted by $\scriptstyle rv$.
\end{quotation}

\medskip
\noindent {\it iii) The inproduct of two vectors}
\smallskip

\noindent
We can also define what is called the inproduct of two vectors
$v$ and $w$, denoted by $<v , w>$, as shown in Fig 4. It is the
real number that is given by the length of vector $v$ multiplied by
the length of vector $w$, multiplied by the cosine of the angle
between the two vectors $v$ and $w$. Hence
\begin{equation}
<v , w> = |v||w|\cos\gamma
\end{equation}
where $\gamma$ is the angle between the vectors $v$ and $w$ (see
Fig 4). By means of this inproduct it is possible to express some
important geometrical properties of space. For example: the inproduct
$<v , w>$ of two non-zero vectors equals zero iff the two vectors are
orthogonal to each other. On the other hand there is also a
straightforward relation between the inproduct of a vector with
itself and the length of this vector
\begin{equation}
<v,v> = |v|^2
\end{equation}

\vskip 0.5 cm

\hskip 3.5 cm \includegraphics{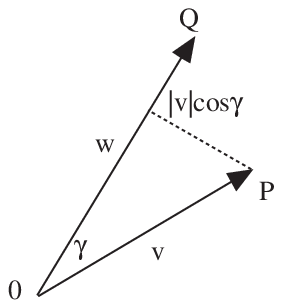}

\begin{quotation}
\noindent \baselineskip= 7pt \smallroman Fig 4 : A representation
of the inproduct of two vectors $\scriptstyle v$ and
$\scriptstyle w$, denoted by $\scriptstyle <v , w>$.
\end{quotation}

\medskip
\noindent {\it iv) An orthonormal base of the vector space}
\smallskip

\noindent
For each finite dimensional vector space with an inproduct it is
possible to define an orthonormal base. For our case of the three
dimensional real vector space that we use to describe the points of
the three dimensional Euclidean space it is a set of three orthogonal
vectors with the length of each vector equal to $1$ (see Fig 5).

\vskip 0.5 cm

\hskip 3 cm \includegraphics{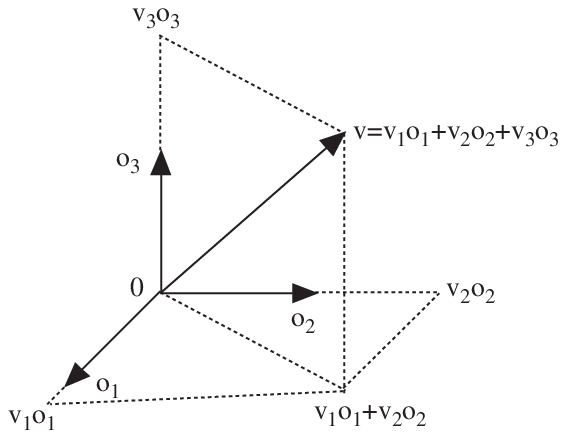}

\begin{quotation}
\noindent \baselineskip= 7pt \smallroman Fig 5 : An orthonormal base
$\scriptstyle \{h_1,h_2,h_3\}$ of the three dimensional real vector
space. We can write an arbitrary vector
$\scriptstyle v$ as a sum of the three base vectors multiplied
respectively by real numbers $\scriptstyle v_1, v_2$ and $\scriptstyle
v_3$. There numbers are called the Cartesian coordinates of the vector
$\scriptstyle v$ for the orthonormal base
$\scriptstyle \{h_1,h_2,h_3\}$.
\end{quotation}

\noindent
Hence the set $\{h_1, h_2, h_3\}$ is an orthonormal base of our
vector space 
$\Bbb{R}^3$ iff
\begin{equation}
\begin{array}{lll}
<h_1,h_1> = 1 & <h_1,h_2> = 0 & <h_1,h_3> = 0 \\
<h_2,h_1> = 0 & <h_2,h_2> = 1 & <h_2,h_3> = 0 \\
<h_3,h_1> = 0 & <h_3,h_2> = 0 & <h_3,h_3> = 1
\end{array}
\end{equation}
We can write each vector $v \in \Bbb{R}^3$ as the sum of
these base vectors respectively multiplied by real numbers $v_1, v_2$
and
$v_3$ (see Fig 5). Hence:
\begin{equation}
v = v_1h_1+v_2h_2+v_3h_3
\end{equation}
The numbers $v_1,v_2$ and $v_3$ are called the Cartesian
coordinates\footnote{It was Ren{\'e} Descartes who introduced
this mathematical representation of our three dimensional Euclidean
space.} of the vector
$v$ for the orthonormal base
$\{h_1,h_2,h_3\}$.

\medskip
\noindent {\it v) The Cartesian representation of space}
\smallskip

\noindent
As we have fixed the origin $0$ of our vector space we can also fix
one specific orthonormal base, for example the base
$\{h_1,h_2,h_3\}$, and decide to express each vector $v$ by means of
the Cartesian coordinates with respect to this fixed base. We will
refer to such a fixed base as a Cartesian base. Instead of
writing $o = v_1h_1+v_2h_2+v_3h_3$ it is common practice to
write $v=(v_1,v_2,v_3)$, only denoting the three Cartesian coordinates
and not the Cartesian base, which is fixed
now anyway. As a logical consequence we denote $h_1 = (1,0,0)$, $h_2=
(0,1,0)$ and
$h_3 = (0,0,1)$. It is an easy exercise to show that the addition,
multiplication with a real number and the inproduct of vectors are
given by the following formulas. Suppose that we consider two vectors
$v = (v_1,v_2,v_3)$ and $w = (w_1,w_2,w_3)$ and a real number $r$,
then we have:
\begin{equation}
\begin{array}{l}
v+w = (v_1+w_1,v_2+w_2,v_3+w_3) \\
rv = (rv_1,rv_2,rv_3) \\
<v,w> = v_1w_1+v_2w_2+v_3w_3
\end{array}
\end{equation}
These are very simple mathematical formulas. The addition of vectors
is just the addition of the Cartesian coordinates of these vectors,
the multiplication with a real number is just the multiplication of the
Cartesian coordinates with this number, and the inproduct of vectors
is just the product of the Cartesian coordinates. This is one of the
reasons why the Cartesian representation of the points of space is
very powerful. 

\medskip
\noindent {\it vi) The representation of space by means of spherical
coordinates}
\smallskip

\noindent
It is possible to introduce many systems of coordination of space. We
will in the following use one of these other systems: the
spherical coordinate system.

\vskip 0.5 cm

\hskip 3.5 cm \includegraphics{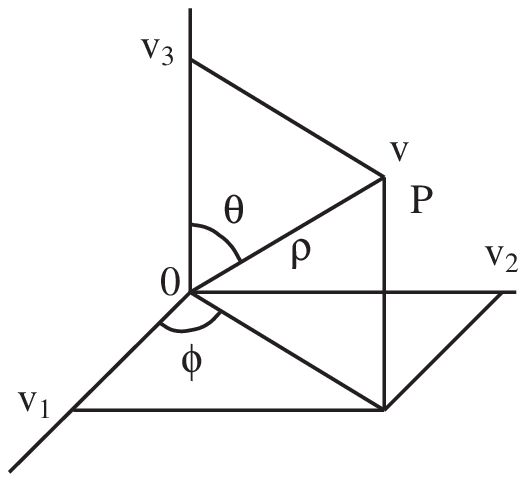}

\begin{quotation}
\noindent \baselineskip= 7pt \smallroman Fig. 6 : A point $\scriptstyle P$ with
cartesian coordinates $\scriptstyle v_1, v_2$ and $\scriptstyle v_3$,
and spherical coordinates $\scriptstyle \rho,
\theta$ and $\scriptstyle \phi$.
\end{quotation}

\noindent
 We show the spherical coordinates
$\rho$, $\theta$ and $\phi$ of a point $P$ with Cartesian
coordinates $(v_1, v_2, v_3)$ in Figure 6. We have the
following well known and easy to verify relations between the two sets
of coordinates (see Fig 6).
\begin{equation}
\begin{array}{l}
v_1 = \rho \sin\theta \cos\phi \\
v_2 = \rho \sin\theta \sin\phi \\
v_3 = \rho \cos\theta
\end{array}
\end{equation}
 
\subsection{The states of the
quantum machine entity}

\noindent
We have introduced in the foregoing section some elementary mathematics
necessary to handle the quantum machine entity. First we will
define the possible states of the entity and then the experiments
we can perform on the entity.

\vskip 0.5 cm

\hskip 2.7 cm \includegraphics{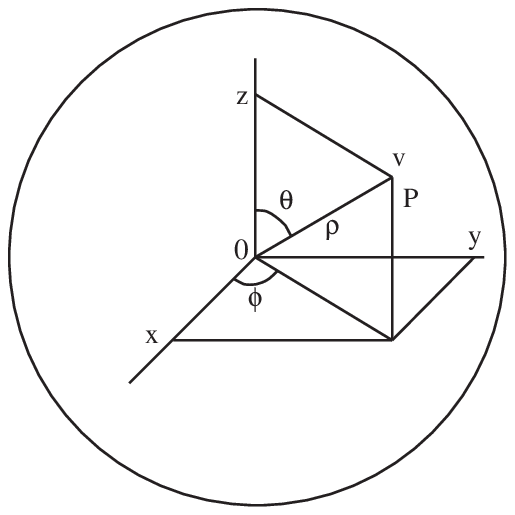}

\begin{quotation}
\noindent \baselineskip= 7pt \smallroman Fig. 7 : The quantum machine entity
$\scriptstyle P$ with its vector representation $\scriptstyle
v$, its cartesian coordinates
$\scriptstyle x, y$ and
$\scriptstyle z$, and its spherical coordinates $\scriptstyle \rho,
\theta$ and $\scriptstyle \phi$.
\end{quotation}

\noindent 
The quantum machine entity is a point
particle
$P$ in three dimensional Euclidean space that we represent by a vector
$v$. The states of the quantum machine entity are the different
possible places where this point particle can be, namely inside or on
the surface of a spherical ball (that we will denote by $ball$) with
radius
$1$ and center
$0$ (Fig 7)\footnote{In the earlier presentations of the quantum
machine [5], [7], [8] and [9], we only considered the points
on the surface of the sphere to be the possible states of the quantum
machine entity. If we want to find a model that is strictly equivalent
to the spin model for the spin${1 \over 2}$ of a quantum entity, this
is what we have to do. We will however see that it is fruitful, in
relation with a possible solution of one of the quantum paradoxes, to
introduce a slightly more general model for the quantum machine and
also allow points of the interior of the sphere to represent states
(see also [6] for the representation of this more general quantum
machine)}.

\par Let us denote
the set of states of the quantum machine entity by
$\Sigma_{cq}$\footnote{The subscript $cq$ stands for `completed
quantum mechanics. We will see indeed in the following that the
interior points of $ball$ do not correspond to vector states of
standard quantum mechanics. If we add them anyhow to the set of
possible states, as we will do here, we present a completed version
of standard quantum mechanics. We will come back to this point in
detail in the following sections.}. We will denote a specific state
corresponding to the point particle being in the place indicated by
the vector $v$ by the symbol
$p_v$. So we have:
\begin{equation}
\Sigma_{cq} = \{p_v\ \vert\ v \in \Bbb{R}^3, |v| \le 1\}
\end{equation}
Let us now explain in which way we
interact, by means of experiments, with this quantum machine entity. As
we have defined the states it would seem that we can `know' these
states just by localizing the point inside the sphere (by means of a
camera and a picture for example, or even just by looking at the
point). This however is not the case. We will define very specific
experiments that are the `only' ones at our disposal to find
out `where' the point is. In this sense it would have been more
appropriate to define first the experiments and afterwards the states
of the quantum machine entity. We will see that this is the way that
we will proceed when we introduce the spin of a spin${1 \over 2}$
quantum entity\footnote{It is essential for the reader to understand
this point. In our model we have defined the states of the quantum
machine entity, but actually there is no camera available to `see'
these states. The only experiments available are the ones that we
will introduce now.}.

\subsection{The experiments of the quantum machine}

\noindent
Let us now introduce the experiments. To do this we consider the point
$u$ and the diametrically opposite point
$-u$ of the surface of the sphere $ball$. We install an elastic
strip (e.g. a rubber band) of 2 units of length, such that it is
fixed with one of its end-points in
$u$ and the other end-point in
$-u$ (Fig 8,a).

\par As we have explained, the state
$p_v$ represents the point particle $P$ located in the point $v$. We
will limit ourselves in this first introduction of the quantum
machine to states $p_v$ where $v = 1$ and hence $P$ is on the surface
of the sphere. Later we will treat the general case. Let us now
describe the experiment. Once the elastic is installed, the particle
$P$ falls from its original place $v$ orthogonally onto
the elastic, and sticks to it (Fig 8,b). Then, the elastic breaks at
some arbitrary point. Consequently the particle
$P$, attached to one of the two pieces of the elastic (Fig 8,c), is
pulled to one of the two end-points $u$ or
$-u$  (Fig 8,d). Now, depending on whether the
particle $P$ arrives in $u$ (as in  Fig 8) or in
$-u$, we give the outcome $o_1$ or $o_2$ to
the experiment. We will denote this experiment by the symbol $e_u$
and the set of experiments connected to the quantum machine by ${\cal
E}_{cq}$. Hence we have:

\begin{equation}
{\cal E}_{cq} = \{e_u \ \vert\ u \in \Bbb{R}^3, |u| = 1\}
\end{equation}

\vskip 0.5 cm

\hskip 2.1 cm \includegraphics{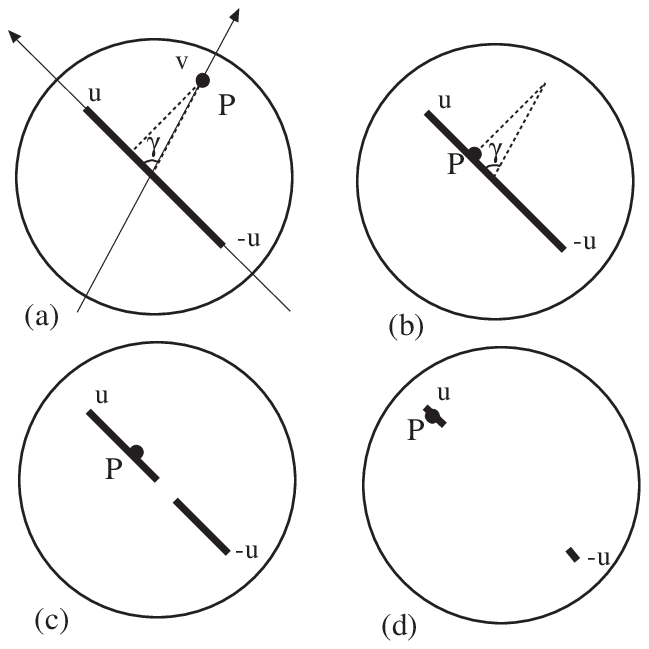}

\begin{quotation}
\noindent \baselineskip= 7pt \smallroman Fig. 8 : A representation of the 
quantum machine. In (a) the particle $\scriptstyle P$ is in
state $\scriptstyle p_v$, and
the elastic corresponding to the experiment $\scriptstyle
e_u$ is installed between the two diametrically opposed
points $\scriptstyle u$ and 
$\scriptstyle -u$. In (b) the particle
$\scriptstyle P$ falls orthogonally onto the elastic  and sticks to
it. In (c) the elastic breaks and the particle $\scriptstyle P$ is
pulled  towards the point
$\scriptstyle u$, such that (d) it arrives at the
point 
$\scriptstyle u$, and the experiment $\scriptstyle
e_u$ gets the outcome
$\scriptstyle  o_1$.
\end{quotation}

\vskip 0.5 cm

\hskip 3.8 cm \includegraphics{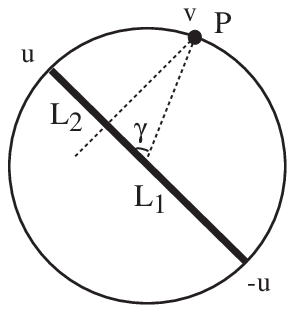}

\begin{quotation}
\noindent \baselineskip= 7pt \smallroman Fig. 9 : A representation of the 
experimental process in the plane where it takes place. The elastic of
length 2, corresponding to the experiment
$\scriptstyle e_u$, is installed between $\scriptstyle
u$ and 
$\scriptstyle -u$. The probability, $\scriptstyle
\mu(e_u, p_v, o_1)$, that the particle 
$\scriptstyle P$ ends up in point
$\scriptstyle u$ under influence of the experiment
$\scriptstyle e_u$ is given by the length of the piece of
elastic
$\scriptstyle L_1$ divided by the total length of the elastic. The
probability,
$\scriptstyle
\mu(e_u,  p_v, o_2)$, that the particle
$\scriptstyle P$ ends up  in point $\scriptstyle
-u$ is given by the length of the piece of
elastic
$\scriptstyle L_2$ divided by the total length of the elastic.
\end{quotation}

\subsection{The transition probabilities of the quantum machine}
\label{secprob}

\noindent
Let us now calculate the probabilities that are involved in these
experiments such that we will be able to show later that
they equal the quantum probabilities connected to the quantum
experiments on a spin${1
\over 2}$ quantum particle.

The probabilities are easily calculated. The probability,
 $\mu(e_u, p_v, o_1)$, that the particle $P$
ends up in point $u$ and hence experiment
$e_u$ gives outcome $o_1$, when the quantum machine entity is in
state $p_v$, is given by the length of the piece of elastic
$L_1$ divided by the total length of the elastic (Fig 9). The
probability,
$\mu(e_u, p_v, o_2)$, that the particle
$P$ ends up in point
$-u$, and hence experiment $e_u$
gives outcome
$o_2$, when the quantum machine entity is in state $p_v$, is given by
the length of the piece of elastic
$L_2$ divided by the total length of the elastic. This gives us:
\begin{eqnarray}
\mu(e_u, p_v, o_1) &=& {L_1\over 2} = {1 \over
2}(1 + \cos\gamma) = \cos^2{\gamma\over 2} \label{transqm01} \\
\mu(e_u, p_v, o_2) &=& {L_2\over 2} = {1 \over
2}(1 - \cos\gamma) = \sin^2{\gamma\over 2} \label{transqm02}
\end{eqnarray}
As we will see in next section, these are exactly the probabilities
related to the spin experiments on the spin of a spin${1 \over 2}$
quantum particle.

\section{The spin of a spin${1 \over 2}$ quantum particle}

\noindent
Let us now describe the spin of a spin${1 \over 2}$ quantum particle
so that we can show that our quantum machine is equivalent to it.
Here we will proceed the other way around and first describe in which
way this spin manifests itself experimentally.

\subsection{The experimental manifestation of the spin}

\noindent
The experiment showing the first time the property of spin
for a quantum particle was the one by Stern and Gerlach [10], and the
experimental apparatus involved is called a
Stern-Gerlach apparatus. It essentially consists of a magnetic field
with a strong gradient, oriented in a particular direction $u$ of
space (Fig 10).

\vskip 0.5 cm

\hskip 1.3 cm \includegraphics{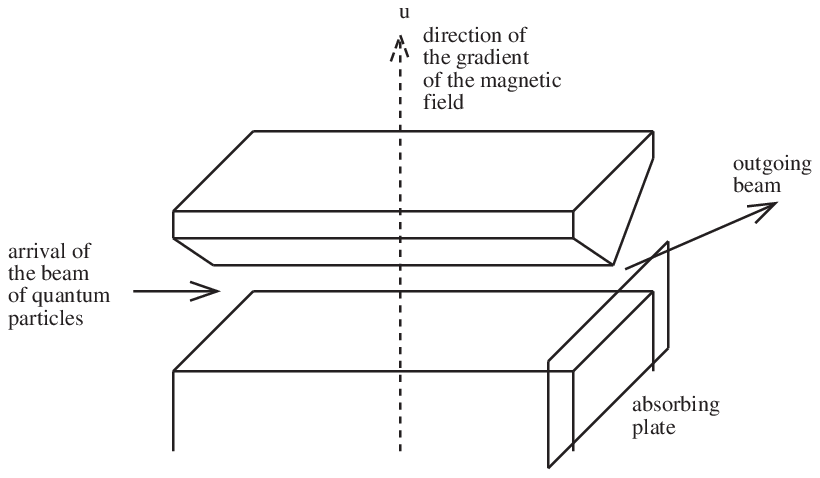}

\begin{quotation}
\noindent \baselineskip= 7pt \smallroman Fig 10: The Stern-Gerlach
apparatus measuring the spin of a spin$\scriptstyle {1 \over 2}$
quantum particle. The particle beam comes in from the left and
passes through a magnetic field with a strong gradient in the
$\scriptstyle u$ direction. The beam is split in
two beams, one upwards, absorbed by a plate, and one downwards,
passing through.
\end{quotation}

\medskip
\noindent
A beam of quantum particles of spin ${1 \over 2}$ is directed
through the magnet following a path orthogonal to the direction of
the gradient of the magnetic field. The magnetic field splits the
beam into two distant beams which is a very unexpected phenomenon if
the situation would be analysed by classical physics. One beam travels
upwards in the direction of the gradient of the magnetic field and one
downwards. The beam directed downwards is absorbed by a plate that
covers this downwards part behind the magnetic field in the regions
where the beams emerge. The upwards beam passes through and
out of the Stern-Gerlach apparatus. We will denote such a Stern-Gerlach
experiment by
$f_u$ where
$u$ refers to the direction of space of the gradient of the magnetic
field. 
\par
If we consider this experiment being performed
on one single quantum particle of spin ${1 \over 2}$, then it has two
possible outcomes: (1) the particle is deflected upwards and passes
through the apparatus, let us denote this outcome
$o_1$, (2) the particle is deflected downwards, and is absorbed by the
plate, lets denote this outcome $o_2$. The set of experiments that we
will consider for this one quantum particle is the set of all
possible Stern-Gerlach experiments, for each direction $u$ of space,
which we denote by
${\cal E}_{sg}$\footnote{The subscript sg stands for Stern-Gerlach.}.
Hence we have:
\begin{equation}
{\cal E}_{sg} = \{f_u \ \vert\ u\ {\rm a\
direction\ in\ space} \}
\end{equation}

\subsection{The spin states of
a spin${1 \over 2}$ quantum particle}

\noindent
Suppose that we have made a Stern-Gerlach experiment as explained
and we follow the beam of particles that emerges. Suppose that we
perform a second Stern-Gerlach experiment on this beam, with the
gradient of the magnetic field in the same direction $u$ as in the
first preparing Stern-Gerlach apparatus. We will then see that now
the emerging beam is not divided into two beams. All particles of the
beam are deviated upwards and none of them is absorbed by the plate.
This means that the first experiment has `prepared' the particles of
the beam in such a way that it can be predicted that they will all be
deviated upwards: in physics we say that the particles have been
prepared in a spin state in direction
$u$. Since we can make such a preparation for each direction of space
$u$ it follows that the spin of a spin${1 \over 2}$ quantum particle
can have a spin state connected to any direction of space. So we will
have to work out a description of this spin state such that each
direction of space corresponds to a spin state. This is exactly what
quantum mechanics does and we will now expose this quantum
description of the spin states. Since we have chosen to denote the
Stern-Gerlach experiments by $f_u$ where $u$ refers to the
direction of the gradient of the magnetic field, we will use the
vector $v$ to indicate the direction of the spin of
the quantum particle. Let us denote such a state by $q_v$.
This state $q_v$ is then connected with the preparation by
means of a Stern-Gerlach apparatus that is put in a direction of
space $v$. A particle that emerges from this Stern-Gerlach apparatus
and is not absorbed by the plate `is' in spin state
$q_v$. Let us denote the set of all spin states by
$\Sigma_{sg}$. So we have:
\begin{equation}
\Sigma_{sg} = \{q_v\ \vert\ v \ {\rm a\
direction\ in\ space} \}
\end{equation}

\subsection{The transition probabilities of the spin}

\noindent
If we now consider the experimental situation of two
Stern-Gerlach apparatuses placed one after the other. The first
Stern-Gerlach apparatus, with gradient of the magnetic field in
direction $v$, and a plate that absorbs the particles that are
deflected down, prepares particles in spin state $q_v$. The second
Stern-Gerlach apparatus is placed in direction of space $u$ and
performs the experiment $f_u$. We can then measure the relative
frequency of particles being deflected ``up" for example. The
statistical limit of this relative frequency is the probability that a
particle will be deflected upwards. The laboratory results are such
that, if we denote by
$\mu(f_u,q_v,o_1)$ the probability that the experiment $f_u$ makes the
particle deflect upwards - let us denote this as outcome $o_1$ - if
the spin state is $q_v$, and by $\mu(f_u,q_v,o_2)$ the probability that
the experiment $f_u$ makes the particle be absorbed by the plate - let
us denote this outcome by $o_2$ - if the spin state is $q_v$, we have:

\begin{eqnarray}
\mu(f_u, q_v, o_1) &=& \cos^2{\gamma\over 2}  \label{trans01} \\
\mu(f_u, q_v, o_2) &=& \sin^2{\gamma\over 2} \label{trans02}
\end{eqnarray}
where $\gamma$ is the angle between the two vectors $u$ and $v$.

\subsection{Equivalence of the quantum machine with the spin}

\noindent
It is obvious that the quantum machine is a model for the spin of a
spin${1 \over 2}$ quantum entity. Indeed we just have to identify the
state $p_v$ of the quantum machine entity with the state $q_v$ of the
spin, and the experiment $e_u$ of the quantum machine with the
experiment $f_u$ of the spin for all directions of space $u$ and $v$.
Then we just have to compare the transition probabilities for the
quantum machine derived in formula's
\ref{transqm01} and \ref{transqm02} with the ones measured in the
laboratory for the Stern-Gerlach experiment\footnote{We do not
consider here the states of the quantum machine corresponding to
points of the inside of the sphere. They do not correspond to states
that we have identified already by means of the Stern-Gerlach
experiment. This problem will be analysed in detail further on.}. Let
us show in the next section that the standard quantum mechanical
calculation leads to the same transition probabilities.

\section{The quantum description of the spin} \label{quantspin}

\noindent
We will now explicitly introduce the quantum mechanical description
of the spin of a spin${1 \over 2}$ quantum particle. As
mentioned in the introduction, quantum mechanics makes a
very abstract use of the vector space structure
 for the description of the states of the quantum
entities. Even the case of the simplest quantum model, the one
we are presenting, the  quantum description is rather abstract
as will become obvious in the following. As we have mentioned
in the introduction, it is the abstract nature of the quantum
description that is partly at the origin of the many conceptual
difficulties concerning quantum mechanics. For this reason we will
introduce again quantum mechanics step by step.

\subsection{The quantum description of the spin states}

\noindent
Quantum mechanics describes the different spin states $q_u$ for
different directions of space
$u$ by the unit vectors of a two
dimensional complex vector space, which we will denote by
$\Bbb{C}^2$. This means that the spin states are not described by
unit vectors in a three dimensional real space, as it is the case for
our quantum machine model, but by unit vectors in a two dimensional
complex space. This switch from a three dimensional real space to a
two dimensional complex space lies at the origin of many of the
mysterious aspects of quantum mechanics. It also touches some
deep mathematical corresondences (the relation between SU(2) and
SO(3) for example) of which the physical content has not yet been
completely unraveled. Before we completely treat the
spin states, let us give an elementary description of
$\Bbb{C}^2$ itself.

\bigskip
\noindent
{\it i) The vector space $\Bbb{C}^2$}
\medskip

\noindent
$\Bbb{C}^2$ is a two-dimensional vector space over the field of the
complex numbers. This means that each vector of
$\Bbb{C}^2$ - in its Cartesian representation - is of the form
$(z_1,z_2)$ where
$z_1$ and
$z_2$ are complex numbers. The addition of vectors and the
multiplication of a vector with a complex number are defined
as follows: for $(z_1, z_2), (t_1, t_2) \in \Bbb{C}^2$ and for $z
\in \Bbb{C}$ we have:
\begin{equation}
\begin{array}{l}
(z_1, z_2) + (t_1, t_2) = (z_1 + t_1, z_2 + t_2) \\
z (z_1, z_2) = (z z_1, z z_2)
\end{array}
\end{equation}
There exists an inproduct on $\Bbb{C}^2$: for
$(z_1, z_2), (t_1, t_2) \in \Bbb{C}^2$ we have:
\begin{equation}
<(z_1, z_2),(t_1, t_2)> = z_1^* t_1 + z_2^* t_2 
\end{equation}
where $*$ is the complex conjugation in $\Bbb{C}$. 
We remark that $< , >$ is linear in the second variable and conjugate
linear in the first variable. More explicitly this means that for
$(z_1, z_2), (t_1, t_2), (r_1,r_2) \in \Bbb{C}^2$ and $z, t \in Bbb{C}$
we have:
\begin{eqnarray}
<z(z_1,z_2) + t (t_1,t_2), (r_1, r_2)> &=& z^*<(z_1,z_2),(r_1,r_2)>
\\ \nonumber
& & + t^*<(t_1,t_2),(r_1,r_2)> \\
<(r_1,r_2), z(z_1,z_2) + t(t_1,t_2)> &=& z<(r_1,r_2),(z_1,z_2)> \\
\nonumber
& & + t<(r_1,r_2),(t_1,t_2)>
\end{eqnarray}

The
inproduct defines the `length' or better `norm' of a vector $(z_1,z_2)
\in \Bbb{C}^2$ as:
\begin{equation}
|(z_1,z_2)| = \sqrt{<(z_1,z_2),(z_1,z_2)>} =
\sqrt{|z_1|^2+|z_2|^2}
\end{equation}
and it also defines an orthogonality relation on the vectors of
$\Bbb{C}^2$: we say that two vectors $(z_1, z_2), (t_1, t_2) \in
\Bbb{C}^2$ are orthogonal and write $(z_1, z_2) \perp (t_1,
t_2)$ iff:
\begin{equation}
 <(z_1, z_2), (t_1,
t_2)> \ = 0
\end{equation}
A unit vector in $\Bbb{C}^2$ is a vector with norm equal to
$1$. Since the spin states are represented by these unit vectors
we introduce a special symbol $U(\Bbb{C}^2)$ to indicate the
set of unit vectors of $\Bbb{C}^2$. Hence:
\begin{equation}
(z_1,z_2) \in U(\Bbb{C}^2) \Leftrightarrow
\sqrt{|z_1|^2+|z_2|^2} = 1
\end{equation}
The vector space has the Cartesian orthonormal base formed by
the vectors
$(1,0)$ and
$(0,1)$ because we can obviously write each vector $(z_1,z_2)$ in the
following way: 
\begin{equation}
(z_1,z_2) = z_1(1,0) + z_2(0,1) 
\end{equation}
\bigskip
\noindent
{\it ii) The quantum representation of the spin states}
\medskip
\noindent
Let us now make explicit the quantum mechanical representation of the
spin states. The state
$q_v$ of the spin of a spin${1 \over 2}$ quantum particle
in direction
$v$ is represented by the unit vector $c_v$ of
$U(\Bbb{C}^2)$:
\begin{equation}
c_v = (\cos{\theta \over 2}  e^{i{\phi
\over 2}}, \sin{\theta \over 2}  e^{-i{\phi \over 2}}) =
c(\theta,\phi)
\label{raystate}
\end{equation}
where $\theta$ and $\phi$ are the spherical coordinates of the
unit vector
$v$. It is easy to verify that these are unit vectors in
$\Bbb{C}^2$. Indeed,
\begin{equation}
|c(\theta,\phi)| = \cos^2{\theta
\over 2} + \sin^2{\theta \over 2} = 1
\end{equation}
It is also easy to verify that two quantum states that correspond to
opposite vectors $v = (1, \theta, \phi)$ and $- v = (1, \pi -
\theta,
\phi+\pi)$ are orthogonal. Indeed, a calculation shows
\begin{equation}
<c_v, c_{-v}> = 0
\end{equation}
This gives us the complete quantum mechanical description of the spin
states of the spin of a spin${1 \over 2}$ quantum particle. 

\subsection{The quantum description of the spin experiments}
We now
have to explain how the spin experiments
are described in the quantum formalism. As we explained already in
detail, a spin experiment $f_u$ in the laboratory is executed by means
of a Stern-Gerlach apparatus. To describe these experiments quantum
mechanically we have to introduce somewhat more advanced
mathematics, although still easy to master for those readers who have
had no problems till now. So we encourage them to keep with us.

\bigskip
\noindent
{\it i) Projection operators}
\medskip

\noindent
The first concept that we have to introduce for the description
of the experiments\footnote{We use `experiment' to indicate the
interaction with the physical entity. A measurement in this sense is a
special type of experiment.} is that of an operator or matrix. We only
have to explain the case we are interested in, namely the
case of the two dimensional complex vector space that describes the
states of the spin of a spin${1 \over 2}$ quantum particle. In this
case an operator or matrix
$H$ consists of four complex numbers
\begin{equation}
H = \left(
\begin{array}{ll}
z_{11} & z_{12} \\
z_{21} & z_{22}
\end{array} \right)
\end{equation}
that have a well defined operation on the vectors of $\Bbb{C}^2$: for
$(z_1, z_2) \in \Bbb{C}^2$ we have
$H(z_1,z_2) = (s_1,s_2)$ where
\begin{equation}
(s_1,s_2) = (z_{11}z_1 + z_{21}z_2,z_{12}z_1 + z_{22}z_2)
\end{equation}
It is easily verified that this operation is linear, which means
that for $(z_1, z_2), (t_1,t_2) \in \Bbb{C}^2$ and $z, t \in
\Bbb{C}$ we have:
\begin{equation}
H(z(z_1,z_2) + t(t_1,t_2)) = zH(z_1,z_2) + tH(t_1,t_2)
\end{equation}
Moreover it can be shown that every linear operation on the vectors of
$\Bbb{C}^2$ is represented by a matrix. There is a unit
operator $I$ and a zero operator $0$ that respectively maps each
vector onto itself and on the zero vector $(0,0)$:
\begin{equation}
I = \left( \begin{array}{ll}
1 & 0 \\
0 & 1
\end{array} \right) \quad 
0 = \left( \begin{array}{ll}
0 & 0 \\
0 & 0
\end{array} \right)
\end{equation}
The sum of two operators $H_1$ and $H_2$ is defined as the operator
$H_1 + H_2$ such that 
\begin{equation}
(H_1 + H_2)(z_1, z_2) = H_1(z_1,z_2) + H_2(z_1,z_2)
\end{equation}
and the product
of these two operators is the operator $H_1
\cdot H_2$ such that
\begin{equation}
(H_1 \cdot H_2)(z_1,z_2) =
H_1(H_2(z_1,z_2))
\end{equation}
We need some additional concepts and also a special
set of operators to be able to explain how experiments are described
in quantum mechanics.

\par  We say
that a vector $(z_1, z_2)
\in \Bbb{C}^2$ is an `eigenvector' of the operator $H$ iff for
some $z \in \Bbb{C}$ we have:
\begin{equation}
H(z_1,z_2) = z(z_1,z_2)
\end{equation}
If $(z_1,z_2)$ is different from $(0,0)$ we say that $z$ is the
eigenvalue of the operator $H$ corresponding to the eigenvector
$(z_1,z_2)$.

Let us introduce now the special type of operators that we need to
explain how experiments are described by quantum mechanics. A
projection operator $P$ is an operator such that $P^2 = P$ and such
that for $(z_1, z_2), (t_1,t_2) \in \Bbb{C}^2$ we have:
\begin{equation} \label{hermi}
<(z_1,z_2), P(t_1,t_2)> \ = \ <P(z_1,z_2), (t_1,t_2)>
\end{equation}
We remark that a projection operator, as we have defined it here, is
sometimes called an orthogonal projection operator. Let us
denote the set of all projection operators
by ${\cal L}(\Bbb{C}^2)$. 
\par If
$P \in {\cal L}(\Bbb{C}^2)$ then also $I
- P \in {\cal L}(\Bbb{C}^2)$. Indeed
$(I - P)(I - P) = I - P - P + P^2 = I - P - P + P = I - P$. 

\par We can show
that the eigenvalues of a projection operator are $0$ or $1$. Indeed,
suppose that $z
\in \Bbb{C}$ is such an eigenvalue of $P$ with eigenvector
$(z_1,z_2) \not= (0,0)$. We then have $P(z_1,z_2) = z(z_1,z_2) =
P^2(z_1,z_2) = z^2(z_1,z_2)$. From this follows that $z^2 = z$ and
hence $z = 1$ or $z = 0$.

\par
Let us now investigate what the possible forms of projection
operators are in our case
$\Bbb{C}^2$. We consider the set $\{P, I-P\}$ for an
arbitrary $P \in {\cal L}(\Bbb{C}^2)$.

\medskip
\noindent
i) Suppose that $P(z_1,z_2) = (0,0)\ \forall\ (z_1,z_2) \in
\Bbb{C}^2$. Then $P = 0$ and $I-P = I$. This means that this
situation is uniquely represented by the set $\{0,I\}$.

\medskip
\noindent
ii) Suppose that there exists an element $(z_1,z_2) \in \Bbb{C}^2$ such 
that $P(z_1,z_2) \not= (0,0)$. Then $P(P(z_1,z_2)) =
P(z_1,z_2)$ which shows that $P(z_1,z_2)$ is an eigenvector of $P$
with eigenvalue $1$. Further we also have that $(I-P)(P(z_1,z_2)) = 0$
which shows that $P(z_1,z_2)$ is an eigenvector of $I-P$ with
eigenvalue $0$. If we further supose that
$(I-P)(y_1,y_2) = 0 \ \forall\ (y_1,y_2) \in \Bbb{C}^2$, then we
have $P = I$ and the situation is represented by the set $\{I,0\}$.

\medskip
\noindent
iii) Let us now suppose that there exists a $(z_1, z_2)$ as in ii)
and an element
$(y_1,y_2)
\in
\Bbb{C}^2$ such that $(I-P)(y_1,y_2) \not= 0$. An analogous
reasoning then shows that $(I-P)(y_1,y_2)$ is an eigenvector of
$I-P$ with eigenvalue $1$ and an eigenvector of $P$ with eigenvalue
$0$. Furthermore we have:
\begin{equation}
\begin{array}{l}
<P(z_1,z_2), (I-P)(y_1,y_2)> \\
 = \ <(z_1,z_2), P(I-P)(y_1,y_2)> \ = \ 0
\end{array}
\end{equation}
which shows that $P(z_1,z_2) \perp (I-P)(y_1,y_2)$. This also proves
that $P(z_1,z_2)$ and $(I-P)(y_1,y_2)$ form an orthogonal base of
$\Bbb{C}^2$, and $P$ is in fact the projection on the one
dimensional subspace generated by $P(z_1,z_2)$, while $I-P$ is the
projection on the one dimensional subspace generated by
$(I-P)(y_1,y_2)$.

\medskip
\noindent These three cases i), ii) and iii) cover all the
possibilities, and we have now introduced all the elements necessary to
explain how experiments are described in quantum mechanics.

\bigskip
\noindent
{\it ii) The quantum representation of the measurements}
\medskip

\noindent
An experiment $e$ in quantum mechanics, with two possible
outcomes $o_1$ and $o_2$, in the case where the states are represented
by the unit vectors of the two dimensional complex Hilbert space
$\Bbb{C}^2$, is represented by such a set
$\{P, I-P\}$ of projection operators satisfying situation
iii). This means that $P \not= 0$ and $I-P \not= 0$. Let us state
now the quantum rule that determines in which way the outcomes occur
and what happens to the state under the influence of a measurement.

\medskip
\noindent
If the state $p$ of the quantum entity is represented by a unit vector
$(z_1,z_2)$ and the experiment $e$ by the set of non zero projection
operators $\{P,I-P\}$, then the outcome $o_1$ of $e$ occurs with
a probability given by 
\begin{equation}
<(z_1,z_2),P(z_1,z_2)>
\end{equation}
and if $o_1$ occurs the
unit vector $(z_1,z_2)$ is changed into the unit vector
\begin{equation}
{P(z_1,z_2)
\over
|P(z_1,z_2)|}
\end{equation}
that represents the new state of the quantum
entity after the experiment $e$ has been performed.
The outcome $o_2$ of $e$ occurs with
a probability given by 
\begin{equation}
<(z_1,z_2),(I-P)(z_1,z_2)>
\end{equation}
and if $o_2$ occurs the
unit vector $(z_1,z_2)$ is changed into the unit vector
\begin{equation}
{(I-P)(z_1,z_2)
\over
|(I-P)(z_1,z_2)|}
\end{equation}
that represents the new state of the quantum
entity after the experiment $e$ has been performed.

\par We remark that we have:
\begin{equation}
\begin{array}{l}
<(z_1,z_2),P(z_1,z_2)> + <(z_1,z_2),(I-P)(z_1,z_2)> \\
= <(z_1,z_2),(z_1,z_2)> = 1
\end{array}
\end{equation}
such that these numbers can indeed serve as probabilities for the
respective outcomes, i.e. they add up to $1$ exhausting all other
possibilities.

\bigskip
\noindent
{\it iii) The quantum representation of the spin experiments}
\medskip

\noindent
Let us now make explicit how quantum mechanics describes the spin
experiments with a Stern-Gerlach apparatus in direction
$u$, hence the experiment $f_u$.

The projection operators that correspond to the spin measurement of
the spin of a spin${1 \over 2}$ quantum
particle in the $u$ direction are given by
$\{P_u, I-P_u\}$ where
\begin{equation} \label{spinproj01}
P_u = {1 \over 2} \left(
\begin{array}{ll}
1 + \cos\alpha & e^{-i\beta}\sin\alpha \\
e^{+i\beta}\sin\alpha & 1 - \cos\alpha
\end{array} \right)
\end{equation}
where $u = (1,\alpha,\beta)$ and hence $\alpha$ and $\beta$ are the
spherical coordinates angles of the vector $u$. We can easily verify
that
\begin{equation} \label{spinproj02}
I-P_u = P_{-u} = {1 \over 2} \left(
\begin{array}{ll}
1 - \cos\alpha & -e^{-i\beta}\sin\alpha \\
-e^{+i\beta}\sin\alpha & 1 + \cos\alpha
\end{array} \right)
\end{equation}
Let us calculate in this case the quantum probabilities. Hence
suppose that we have the spin in a state $q_v$ and a spin
measurement $f_u$ in the direction $u$ is
performed. The quantum rule for the probabilities then gives us the
following result. The probability $\mu(f_u, q_v,
o_1)$ that the experiment
$f_u$ gives an outcome $o_1$ if the quantum particle is in
state
$q_v$ is given by:
\begin{equation} \label{probquant01}
\mu(f_u, q_v, o_1) = <c_v,
P_uc_v>
\end{equation}
And the probability $\mu(f_u, q_v,
o_2)$ that the experiment
$f_u$ gives an outcome $o_2$ if the quantum particle is in
state
$q_v$ is given by:
\begin{equation} \label{probquant02}
\mu(f_u, q_v, o_2) = <c_v,
(I-P_u)c_v>
\end{equation}
This is not a complicated, but a somewhat long calculation. If we
introduce the angle $\gamma$ between the two directions $u$ and
$v$ (see Fig. 11), it is possible to show that the
probabilities can be expressed by means of this angle $\gamma$ in a
simple form.

\vskip 0.5 cm

\hskip 2.7 cm \includegraphics{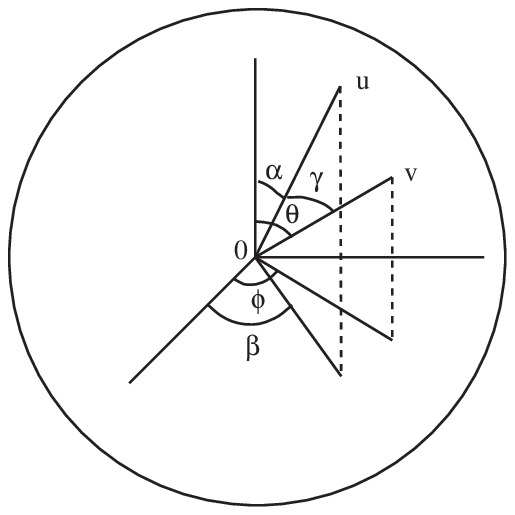}

\begin{quotation}
\noindent \baselineskip= 7pt \smallroman Fig. 11 : Two points
$\scriptstyle v = (1,\theta,\phi)$ and $\scriptstyle
u = (1,\alpha,\beta)$ on the unit sphere and the angle $\gamma$
between the two directions $\scriptstyle (\theta,\phi)$ and
$\scriptstyle (\alpha,\beta)$.
\end{quotation}

\medskip
\noindent
We have:
\begin{equation}
\mu(f_u, q_v, o_1) = \cos^2{\gamma \over 2}
\end{equation}
\begin{equation}
\mu(f_u, q_v, o_2) = \sin^2{\gamma \over 2}
\end{equation}
These are indeed the transition probabilities for the spin
measurement of the spin of a spin${1 \over 2}$ quantum particle found
in the laboratory (see formulas~\ref{trans01},~\ref{trans02}).

\par To see this immediately we can,
without loss of generality, choose the situation of the spin
measurement for the $z$ axis direction. Hence in this case we have $u
= (1,\alpha,\beta)$ with $\alpha = 0$ and $\beta = 0$ for the
experiment giving outcome $o_1$ and $\alpha =
\pi$ and
$\beta = 0$ for the experiment giving the outcome $o_2$. This gives us
for
$f_u$ the projection operators
$\{P_u,P_{-u}\}$ given by:
\begin{equation}
P_u = \left(
\begin{array}{ll}
1 & 0 \\
0 & 0
\end{array} \right) \quad
P_{-u} = \left(
\begin{array}{ll}
0 & 0 \\
0 & 1
\end{array} \right)
\end{equation}
For the probabilities we have:
\begin{equation}
\mu(f_u, q_v, o_1) = <c_v,
P_uc_v> = \cos^2{\theta \over 2}
\end{equation}
\begin{equation}
\mu(f_u, q_v, o_2) = <c_v,
P_{-u}c_v> = \sin^2{\theta \over 2}
\end{equation}

\medskip
\noindent
We mentioned already that
quantum mechanics describes the experiments in a very abstract way.
This is one of the reasons why it is so difficult to understand many
of the aspects of quantum mechanics. Since the quantum machine that we
have presented also generates an isomorphic structure we have shown
that, within our common reality, it is possible to realize this
structure. The quantum machine is a model for the spin of a spin${1
\over 2}$ quantum particle and it is also a model for the abstract
quantum description in the two dimensional complex vector space. This
result will make it possible for us to analyse the hard quantum
problems and to propose new solutions to these problems.

\section{Quantum probability}

\noindent
Probability, as it was first and formally introduced by Laplace and
later axiomatised by Kolmogorov, is a description of our lack of
knowledge about what really happens. Suppose that we say for a
dice we throw, we have one chance out of six that the number $2$ will
be up when it lands on the floor. What is the meaning of this
statement? It is very well possible that the motion of the dice
through space after it has been thrown is completely deterministic and
hence that if we knew all the details we could predict the
outcome with certainty\footnote{We do not state here that complete
determinism is the case, but just want to point out that it could
be.}. So the emergence of the probability ${1 \over 6}$ in the case of
the dice does not refer to an indeterminism in the reality of what
happens with the dice. It is just a description of our lack of
knowledge about what precisely happens. Since we do not know what
really happens during the trajectory of the dice and its landing and
since we also do not know this for all repeated experiments we
will make, for reasons of symmetry, there is an equal chance for each
side to be up after the dice stopped moving. Probability of ${1 \over
6}$ for each event expresses this type of lack of knowledge.
Philosophers say in this case that the probability is `epistemic'.
Probability that finds its origin in nature itself and not in our
lack of knowledge about what really happens is called `ontologic'.

Before the
advent of quantum mechanics there was no important debate about the
nature of the probabilities encountered in the physical world. It was
commonly accepted that probabilities are epistemic and hence can be
explained as due to our lack of knowledge about what really
happens. It is this type of probability - the one encountered in
classical physics - that was axiomatized by Kolmogorov and therefore
it is commonly believed that Kolmogorovian probabilities are
epistemic.

It was rather a shock when physicists found out that the structure
of the probability model that is encountered in quantum mechanics
does not satisfy the axioms of Kolmogorov. Would this mean that
quantum probabilities are ontologic? Anyhow it could only mean
two things: (1) the quantum probabilities are ontologic, or (2) the
axiomatic system of Kolmogorov does not describe all types of
epistemic probabilities. In this second case quantum probabilities
could still be epistemic.

\subsection{Hidden variable theories}

\noindent
Most of investigations that physicist, mathematicians and
philosophers carried out regarding the nature of the quantum
probabilities pointed in the direction of ontologic
probabilities. Indeed, a whole area of research was born
specifically with the aim of investigating this problem, it was called
`hidden variable theory research'. The reason for this naming is the
following: if the quantum probabilities are merely epistemic, it must
be possible to build models with `hidden variables', their prior
absence causing the probabilistic description. These models must be
able to substitute the quantum models, i.e. they are equivalent, and
they entail explicitly epistemic probabilities, in the sense that a
randomisation over the hidden variables gives rise to the quantum
probabilities.

Actually physicists trying to construct such hidden
variable models had thermodynamics in mind. The
theory of thermodynamics is independent of classical mechanics, and
has its own set of physical quantities such as pressure, volume,
temperature, energy and entropy and its own set of states. It was,
however, possible to introduce an underlying theory of classical
mechanics. To do so, one assumes that every thermodynamic entity
consists of a large number of molecules and the real pure state of the
entity is determined by the positions and momenta of all these
molecules. A thermodynamic state of the entity is then a mixture
or mixed state of the underlying theory. The programme had been found
feasible and it was a great success to derive the laws of
thermodynamics in this way from Newtonian mechanics.

Is it possible to do something similar for quantum mechanics? Is it
possible to introduce extra variables into quantum mechanics such
that these variables define new states, and the description of
the entity based on these new states
is classical? Moreover would quantum mechanics be the
statistical theory that results by averaging over these variables?
This is what scientists working on hidden variable theories were
looking for.

John Von Neumann gave the first proof of the impossibility of hidden
variables for quantum mechanics [1]. One of the
assumptions that Von Neumann made in his proof is that the
expectation value of a linear combination of physical quantities is
the linear combination of the expectation values of the physical
quantities. As remarked by John Bell [11], this assumption is
not justified for non compatible physical quantities, such that,
indeed, Von Neumann's proof cannot be considered conclusive.
Bell constructs in the same reference a hidden variable model for the
spin of a spin${1 \over 2}$ quantum particle, and shows that indeed
Von Neumann's assumption is not satisfied in his model. Bell also
criticizes in his paper two other proofs of the nonexistence of
hidden variables, namely the proof by Jauch and Piron
[12] and the proof by Gleason [13]. Bell
correctly points out the danger of demanding extra assumptions to be
satisfied without knowing exactly what these assumptions mean
physically. The extra mathematical assumptions, criticized by Bell,
were introduced in all these approaches to express the physical idea
that it must be possible to find, in the hidden variable description,
the original physical quantities and their basic algebra. This
physical idea was most delicately expressed, without extra
mathematical assumptions, and used in the impossibility proof of
Kochen and Specker [14]. Gudder [15]
gave an impossibility proof along the same lines as the one of Jauch
and Piron, but now carefully avoiding the assumptions criticized by
Bell. 

One could conclude by stating that every one of these impossibility
proofs consists of showing that a hidden variable theory gives rise
to a certain mathematical structure for the physical quantities
(see [1], [13] and [14]) or for the properties (see [12] and [15]) of
the physical entity under consideration. The physical quantities and
the properties of a quantum mechanical entity do not fit into this
structure and therefore it is impossible to replace quantum mechanics
by a hidden variable theory.

More recently this structural difference between classical entities
and quantum entities has been studied by Accardi within the
category of the probability models itself [16], [17]. Accardi
explicitly defines the concept of Kolmogorovian probability model
starting from the concept of conditional probability. He identifies
Bayes axiom as the one that, if it is satisfied, renders the
probability model Kolmogorovian, i.e. classical. Again this approach of
probability  shows the fundamental difference between a classical
theory and a quantum theory.

A lot of physicists, once aware of this fundamental structural
difference between classical and quantum theories, gave up hope that
it would ever be possible to replace quantum mechanics by a hidden
variable theory; and we admit to have been among them. Meanwhile it
has become clear that the state of affairs is more complicated.

\subsection{Hidden measurement theories}

\noindent
Years ago we managed to built a macroscopic classical entity that
violates the Bell inequalities [18], [19] and [20]. About the same
time Accardi had shown that Bell inequalities are equivalent to his
inequalities characterizing a Kolmogorovian probability model. This
meant that the example we had constructed to violate Bell
inequalities should also violate Accardi's inequalities characterizing
a Kolmogorovian probability model, which indeed proved to be
the case. But this meant that we had given an example of a
macroscopic `classical' entity having a non Kolmogorovian probability
model. This was very amazing and the classification made by many
physicists of a micro world described by quantum mechanics and a macro
world described by classical physics was challenged.
The macroscopic entity with a non Kolmogorovian probability model was
first published in [7] and refined in [8], but essentially it is the
quantum machine that we have presented again in this paper.

We were able to show at that time the state of affairs to be the
following. If we have a physical entity $S$ and we have a lack of
knowledge about the state $p$ of the physical entity $S$, then a theory
describing this situation is necessarily a classical statistical
theory with a Kolmogorovian probability model. If we have a physical
entity $S$ and a lack of knowledge about the measurement $e$ to be
performed on the physical entity $S$, and to be changing the state of
the entity $S$, then we cannot describe this situation by a classical
statistical theory, because the probability model that arises is
non Kolmogorovian. Hence, lack of knowledge about measurements, that
change the state of the entity under study, gives rise to a non
Kolmogorovian probability model. What do we mean by `lack of knowledge
about the measurement $e$'? Well, we mean that the measurement $e$ is
in fact not a `pure' measurement, in the sense that there are hidden
properties of the measurement, such that the performance of $e$
introduces the performance of a `hidden' measurement, denoted
$e^\lambda$, with the same set of outcomes as the measurement $e$. The
measurement
$e$ consists then in fact of choosing one way or another between one
of the hidden measurements
$e^\lambda$ and then performing the chosen measurement $e^\lambda$.

We can very easily see how these hidden measurements appear
in the quantum machine. Indeed, if we call the measurement
$e_u^\lambda$ the measurement that consists in performing $e_u$ and
such that the elastic breaks in the point $\lambda$ for some $\lambda
\in [-u,u]$, then, each time $e_u$ is performed, it is actually one of
the $e_u^\lambda$ that takes place. We do not control this, in the
sense that the $e_u^\lambda$ are really `hidden measurements' that we
cannot choose to perform. The probability $\mu(e_u, p_v, o_1)$ that
the experiment $e_u$ gives the outcome $o_1$ if the entity is in
state $p_v$ is a randomasation over the different situations where
the hidden measurements $e^\lambda_u$ gives the outcome $o_1$ with
the entity in state $p_v$. This is exactly the way we have calculated
this probability in section~\ref{secprob}.

\subsection{Explaining the quantum probabilities}
 
\noindent
First of all we have to mention that it is possible to construct a
`quantum machine' model for an arbitrary quantum mechanical
entity. This means that the `hidden measurement' explanation can
be adopted generally, explaining also the origin of the quantum
probabilities for a general quantum entity. We refer to [8], [21],
[22], [23] and [24] for a demonstration of the fact that every
quantum mechanical entity can be represented by a hidden measurement
model. What is now the consequence of this for the explanation
of the quantum probabilities? It proves that quantum probability is
epistemic, and hence conceptually no ontologic probability has to be
introduced to account for it. The quantum
probability however finds its origin in the lack of knowledge that we
have about the interaction between the measurement and the physical
entity. In this sense it is a type of probability that is
non-classical and does not appear in the classical statistical
theories. Quantum mechanical states are pure states and descriptions
of the reality of the quantum entity. This means that the `hidden
variable' theories which try to make quantum theory into a classical
statistical theory are doomed to fail, as the mentioned no-go theorems
for hidden variable theories had proven already without explaining
why. Our approach allows to understand why. Indeed, there is another
type of epistemic probability than the one identified in the classical
statistical theories, namely the probability due to a lack of
knowledge on the interaction between the measurement and the physical
entity under study. It is natural that this new type of epistemic
probability cannot be eliminated from a theory that describes the
reality of the physical entity, because it appears when one
incorporates the experiments related to the measurements of the
properties of the physical entity in question. Therefore it is also
natural that it remains present in the physical theory describing the
reality of the physical entity, but it has no ontologic nature. This
means that, with the explanation of the quantum probabilities put
forward here, quantum mechanics does not contradict determinism for
the whole of reality\footnote{Our explanation does
of course not prove that the whole of reality is deterministic. It
shows that quantum mechanics does not give us an argument for the
contrary.}.

\subsection{Quantum, classical and intermediate} \label{intermediate}

\noindent
If we demand for the quantum machine that the elastic can break at
everyone of its points, and the breaking of a piece is such that it
is  proportional to the length of this piece, then this hypothesis
fixes the possible `amount' of lack of knowledge about the
interaction between the  experimental apparatus and the physical
entity. Indeed, only certain type of elastic can be used to perform
the  experiments. On the other hand, we can easily imagine elastic that
break according to different laws depending on their physical 
construction. Let us introduce the following different kinds of
elastics: at one extremity we consider elastics that can break in
everyone of its points and such that the breaking of a piece is
proportional to the length of this piece. These are the ones we
have already considered, and  since they lead to a pure quantum
structure, we call them quantum elastics. At the other extremity, we
consider a type of elastic that can only break in one point, and let
us suppose, for the sake of simplicity, that this point is the middle
of the elastic (in [5], [25], [26] and [27] the general situation is
treated).  This kind of elastic is far from elastic, but since it is
an extreme type of real elastics, we stillgive it that name. We shall
show that if experiments are performed with this class of elastic, the
resulting structures are classical, and therefore we will call them
classical elastics. For the general case, we want to consider a class
of elastics that can only break in a segment of length
$2\epsilon$ around the center of the elastic. Let us call these
$\epsilon$-elastics.

\vskip 0.5 cm

\hskip 3.5 cm \includegraphics{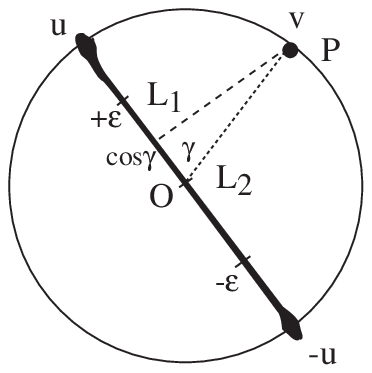}

\begin{quotation}
\noindent \baselineskip= 7pt \smallroman Fig 12 : An experiment with
an $\scriptstyle \epsilon$-elastic. The elastic can only break
between the points
$\scriptstyle -\epsilon$ and $\scriptstyle +\epsilon$. $\scriptstyle
L_1$ is the length of the interval where the elastic can break such
that the point $\scriptstyle P$ finally arrives in $\scriptstyle u$,
and
$\scriptstyle L_2$ is the length of the interval where the elastic can
break such that the point $\scriptstyle P$ finally arrives at
$\scriptstyle -u$.
\end{quotation}

\medskip
\noindent
The elastic with $\epsilon=0$, hence the
$0$-elastic, is the classical elastic, and the elastic with 
$\epsilon=1$, hence the $1$-elastic, is the quantum elastic. In this
way, the parameter 
$\epsilon$ can be interpreted as representing the magnitude of the
lack of knowledge about the interaction between the measuring
apparatus and the physical entity. If
$\epsilon=0$, and for the experiment
$e_u$ only classical elastics are used, there is no
lack of knowledge, in the sense that all elastics will break at the
same  point and have the same influence on the changing of the state
of the entity. The experiment $e_u$ is then a pure experiment. If
$\epsilon = 1$, and for the experiment $e_u$ only
quantum elastics are used, the lack of knowledge is  maximal, because
the chosen elastic can break at any of its points. In Fig 12 we have
represented a typical situation of an experiment with an
$\epsilon$-elastic, where the elastic can only break between the
points $-\epsilon$ and $+\epsilon$. 

Let us calculate the probabilities $\mu(e_u, p_v, o_1)$ and
$\mu(e_u, p_v, o_2)$ for state-transitions from the state
$p_v$ of the particle $P$  before the experiment $e_u$ to one of the
states $p_u$ or $p_{-u}$. Different cases are possible: 

\medskip
\noindent
(1) If the projection of the point $P$ lies between $-u$ and
$-\epsilon$ (see Fig 12), then
\begin{equation}
\begin{array}{ll}
\mu(e_u, p_v, o_1) = 0 & \mu(e_u, p_v, o_2) = 1
\end{array}
\end{equation}
(2) If the projection of the point $P$ lies between $+\epsilon$ and
$u$, then
\begin{equation}
\begin{array}{ll}
\mu(e_u, p_v, o_1) = 1 & \mu(e_u, p_v, o_2) = 0
\end{array}
\end{equation}
(3) If the projection of the point $P$ lies between $-\epsilon$ and
$+\epsilon$ then
\begin{eqnarray}
\mu(e_u, p_v, o_1) &=& {1 \over 2\epsilon}(\epsilon - \cos\gamma) \\
\mu(e_u, p_v, o_2) &=& {1 \over 2\epsilon}(\epsilon + \cos\gamma)
\end{eqnarray}
The entity that we describe here is neither quantum, nor
classical, but intermediate. If we introduce these intermediate
entities, then it becomes possible to describe a continuous transition
from quantum to classical (see [5], [25], [26] and [27] for details).
It gives us a way to introduce a specific solution to
`classical limit problem'.

\section{Quantum axiomatics: the operational part}

\noindent
In the foregoing example of the intermediate situation we have the
feeling that we consider a situation that will not fit into standard
quantum mechanics. However the situation is either not classical. But
how could we prove this? This could only be done if we
had an axiomatic formulation of quantum mechanics
and classical mechanics, such that the axioms could be verified on
real physical examples of entities to see whether a certain situation
is quantum or classical or neither. This means that the axioms
have to be formulated by means of concepts that can be identified
properly if a real physical entity is given. This is certainly not
the case for standard quantum mechanics, but within the quantum
structures research large parts of such an axiomatic system has
been realised through the years.

\subsection{State property spaces: the ontological part}

\noindent
By lack of space we can not expose all
the details of an operational axiomatic formulation, but we will
consider the most important ingredients in some detail and
consider the spin of a spin${1 \over 2}$ quantum particle or the
quantum machine as an example.
In the first place we have to formalize the basic concepts: states
and properties of a physical entity. 

\par
\medskip
\noindent {\it i) The states of the entity $S$}

\par
\medskip
\noindent
With each entity $S$ corresponds a well defined set of states
$\Sigma$ of the entity. This are the modes of being of the entity.
This means that at each moment the entity $S$ `is' in a specific
state $p \in \Sigma$.

\medskip
\noindent {\it ii) The properties of the entity $S$}

\par
\medskip
\noindent
Historically quantum axiomatics has been
elaborated mainly by considering the set of properties\footnote{We
have to remark that in the original paper of Birkhoff and Von
Neumann [2], the concept of `operational proposition' is introduced
as the basic concept. An operational proposition is not the same as a
property (see [28], [29]), but it points at the same
structural part of the quantum axiomatic.}. With each entity $S$
corresponds a well defined set of properties ${\cal L}$. The entity
$S$ `has' a certain property or does not have it.
We will respectively say that the property $a \in {\cal L}$ is
`actual' or is `potential' for the entity $S$. 

\par
\medskip
\noindent
To be able to present the axiomatisation of the set of states and
the set of properties of an entity $S$ in a mathematical way, we
have to introduce some additional concepts.
\par
Suppose that the entity $S$ is in a specific state $p \in \Sigma$.
Then some of the properties of $S$ are actual and some are
not (potential). This means that with each state $p \in
\Sigma$ corresponds a set of actual properties, subset of ${\cal L}$.
Mathematically this defines a function $\xi : \Sigma \rightarrow {\cal
P}({\cal L})$, which makes each state $p \in \Sigma$
correspond to the set $\xi(p)$ of properties that are actual in this
state. With the notation
${\cal P}({\cal L})$ we mean the `powerset' of ${\cal L}$, i.e. the
set of all subsets of ${\cal L}$. From now on - and this is
methodologically a step towards mathematical axiomatization - we can
replace the statement `property $a \in {\cal L}$ is actual for the
entity $S$ in state $p \in \Sigma$' by `$a \in \xi(p)$', which is
just an expression of set theory.
\par
Suppose now that for the entity $S$ a specific
property $a \in {\cal L}$ is actual. Then this entity is in a certain
state $p \in \Sigma$ that makes $a$ actual. With each
property
$a
\in {\cal L}$ we can associate the set of states that make this
property actual, i.e. a subset of $\Sigma$. Mathematically this
defines a function $\kappa : {\cal L} \rightarrow {\cal P}(\Sigma)$,
which makes  each property $a \in {\cal L}$ correspond to the set of
states
$\kappa(a)$ that make this property actual. This is a similar step to
axiomatization. Indeed, this time we can replace the statement
`property $a \in {\cal L}$ is actual if the entity $S$ is in state $p
\in \Sigma$' by the set theoretical expression `$p \in \kappa(a)$'. 
\par Summarising the foregoing we now have:
\begin{equation}
\begin{array}{l}
{\rm property\ } a \in {\cal L}\ {\rm is\ actual\ for\ the\
entity\ } S\ {\rm in\ state\ } p \in \Sigma \\ 
\Leftrightarrow a
\in \xi(p) \\
\Leftrightarrow p \in \kappa(a)
\end{array}
\end{equation}
This expresses a fundamental `duality' among states and
properties. We will introduce a
specific mathematical structure to represent an entity
$S$, its states and its properties, taking into account this duality.
We need the set
$\Sigma$, the set
${\cal L}$, and the two functions
$\xi$ and $\kappa$.

\begin{definition} [state property space]
Consider two sets $\Sigma$ and ${\cal L}$ and two functions
\begin{equation}
\begin{array}{ll}
\xi : \Sigma \leftarrow {\cal P}({\cal L}) &
p \mapsto \xi(p) \\
\kappa : {\cal L} \rightarrow {\cal P}(\Sigma) & a \mapsto \kappa(a)
\end{array}
\end{equation}
If $p \in \Sigma$ and $a \in {\cal L}$ we have:
\begin{equation} \label{statprop}
a \in \xi(p) \Leftrightarrow p \in \kappa(a)
\end{equation}
then we say that $(\Sigma, {\cal L}, \xi, \kappa)$ is a state
property space.
The elements of $\Sigma$ are interpreted as states 
and the elements of ${\cal L}$ as properties of the entity $S$. The
interpretation of (\ref{statprop}) is `property
$a$ is actual if
$S$ is in state $p$' \footnote{We remark that it is enough to
give two sets $\Sigma$ and ${\cal L}$ and a function $\xi : \Sigma
\rightarrow {\cal P}({\cal L})$ to define a state property space.
Indeed, if we define the function $\kappa : {\cal L} \rightarrow {\cal
P}(\Sigma)$ such that $\kappa(a) = \{p\ \vert\ p \in \Sigma, a \in
\xi(p)\}$ then $(\Sigma, {\cal L}, \xi, \kappa)$ is a state property
space. This explains why we do not explicitly consider the function
$\kappa$ in the formal approach outlined in [6], [30] and [31] in the
definition of a state property system, which is a specific type of
state property space. Similarly it would be enough to give $\Sigma$,
${\cal L}$ and $\kappa: {\cal L} \rightarrow {\cal P}(\Sigma)$.}
\end{definition}
There are two natural `implication relations' we can introduce
on a state property space. If the situation is such that if `$a \in
{\cal L}$ is actual for
$S$ in state
$p \in \Sigma$' implies that `$b \in {\cal L}$ is actual for $S$ in
state $p \in \Sigma$' we say that the property $a$ implies the
property $b$. This `property implication' relation is expressed by a
mathematical relation on the set of properties (see following
definition). If the situation is such that `$a \in
{\cal L}$ is actual for
$S$ in state
$q \in \Sigma$' implies that `$a \in {\cal}$ is actual for $S$ in
state $p \in \Sigma$' we say that the state $p$ implies the
state $q$. Again we will express this `state implication' by means
of a mathematical relation on the set of states. 

\begin{definition} [state implication and property implication]
Consider a state property space $(\Sigma, {\cal L},
\xi, \kappa)$. For $a, b \in {\cal L}$ we introduce:
\begin{equation} \label{ordprop}
a \prec b \Leftrightarrow \kappa(a) \subset \kappa(b)
\end{equation}
and we say that $a$ `implies' $b$. For $p, q \in \Sigma$ we introduce:
\begin{equation} \label{ordstat}
p \prec q \Leftrightarrow \xi(q) \subset \xi(p)
\end{equation}
and we say that $p$ `implies' $q$ \footnote{Remark that the state
implication and property implication are not defined in a
completely analogous way. Indeed, then we should for
example have written $p
\prec q \Leftrightarrow \xi(p) \subset \xi(q)$. That we have chosen
to define the state implication the other way around is because
historically this is how intuitively is thought about
states implying one another.}.
\end{definition}
We will introduce now the mathematical concept of a pre-order
relation.
\begin{definition} [pre-order relation]
Suppose that we have a set $Z$. We say that $\prec$ is a pre-order
relation on $Z$ iff for $x, y, z \in Z$ we have:
\begin{equation}
\begin{array}{l}
x \prec x \\
x \prec y \ {\rm and}\ y \prec z \Rightarrow x \prec z
\end{array}
\end{equation}
For two elements $x , y \in Z$ such that $x \prec y$ and $y \prec x$
we denote $x \approx y$ and we say that $x$ is equivalent to $y$.
\end{definition}
It is easy to verify that the implication relations that we
have introduced are pre-order relations.
\begin{theorem}
Consider a state property space $(\Sigma, {\cal L}, \xi, \kappa)$,
then $\Sigma, \prec$ and ${\cal L}, \prec$ are pre-ordered sets.
\end{theorem}
We can show the following for a state property space
\begin{theorem}
Consider a state property space $(\Sigma, {\cal L}, \xi, \kappa)$.
 (1) Suppose that $a, b \in {\cal L}$ and $p \in \Sigma$. If $a \in
\xi(p)$ and $a \prec b$, then $b \in \xi(p)$. (2) Suppose that $p, q
\in
\Sigma$ and $a \in {\cal L}$. If $q \in \kappa(a)$ and $p \prec q$
then $p \in \kappa(a)$.
\end{theorem} \label{statprop04}
Proof: (1) We have $p \in \kappa(a)$ and $\kappa(a) \subset \kappa(b)$.
This proves that $p \in \kappa(b)$ and hence $b \in \xi(p)$. (2) We
have
 $a \in \xi(q)$ and $\xi(q) \subset \xi(p)$ and hence $a \in \xi(p)$.
This shows that
$p
\in \kappa(a)$.

\par
\medskip
\noindent
The reader will now better understand why the original studies of
the axiomatization of quantum mechanics have been called quantum
logic. Indeed, we have also used the name `implication'. We will see
that we can also introduce concepts that are close to `disjunction'
and `conjunction'. But we point out again that we are structuring
more than just the logical aspects of entities. We aim at a
formalization of the complete ontological structure of
physical entities.

\subsection{Meet properties and join states}

\noindent
If we have a structure with implications and we are inspired by
logic, we are tempted to wonder about
conjunctions and disjunctions. Here again it becomes clear
that we are studying a quite different situation than the one
analyzed by traditional logic. 

\par Suppose we consider a set of
properties $(a_i)_i$. It is very well possible that there exist
states of the entity $S$ in which all the properties $a_i$ are
actual. This is in fact always the case if $\cap_i\kappa(a_i) \not=
\emptyset$. Indeed, if we consider $p \in \cap_i\kappa(a_i)$ and $S$
in state $p$, then all the properties $a_i$ are actual. If it is
such that the situation where all properties
$a_i$ of a set $(a_i)_i$ and no other are actual is again a property
of the entity $S$, we will denote this new property by $\wedge_ia_i$,
and call it the `meet property' of all $a_i$. Clearly we have
$\wedge_ia_i$ is actual for $S$ in state $p \in \Sigma$ iff $a_i$ is
actual for all $i$ for $S$ in state $p$. This means that we have
$\wedge_ia_i \in \xi(p)$ iff $a_i \in \xi(p) \ \forall i$. This
formulation of the `meet property' gives us the clue how to introduce
it formally in a state property space.

\par 
Suppose now that we consider a set of states $(p_j)_j$ of the entity
$S$. It is very well possible that there exist properties of the
entity such that these properties are actual if $S$ is in any one of
the states $p_j$. This is in fact always the case if $\cap_j\xi(p_j)
\not=
\emptyset$. Indeed suppose that $a \in \cap_j\xi(p_j)$. Then we have
that $a \in \xi(p_j)$ for each one of the states $p_j$, which means
that $a$ is actual if $S$ is in any one of the states $p_j$. If it
is such that the situation where $S$ is in any one of the states
$p_j$ is again a state of $S$, we will denote this new state by
$\vee_jp_j$ and call it the `join state' of all $p_j$. Clearly we
have that a property $a \in {\cal L}$ is actual for $S$ in
state $\vee_jp_j$ iff this property $a$ is actual for $S$ in any of
the states $p_j$. This formulation of the `join state' indicates
again the way we have to introduce it formally in a state property
space\footnote{We remark that we could also try to introduce join
properties and meet states. It is however a subtle, but deep,
property of reality, that this cannot be done on the same level. We
will understand this better when we introduce in the next section the
operational aspects of the axiomatic approach. We will see there that
only meet properties and join states can be operationally defined in
the general situation.}. 
The existence of meet properties and join states will give
additional structure to $\Sigma$ and ${\cal L}$.

\begin{definition} [complete state property space] \label{comstatprop}
Consider a state property space $(\Sigma, {\cal L}, \xi, \kappa)$. We
say that the state property space is `property complete' iff for an
arbitrary set $(a_i)_i, a_i \in {\cal L}$ of properties there exists
a property
$\wedge_ia_i \in {\cal L}$ such that for an
arbitrary state $p \in \Sigma$:
\begin{equation} \label{meetprop}
\wedge_ia_i \in \xi(p) \Leftrightarrow a_i \in \xi(p)\ \forall\ i
\end{equation}
We say that a state property space is `state complete' iff for an
arbitrary set of states $(p_j)_j, p_j \in \Sigma$ there exists a state
$\vee_jp_j \in \Sigma$ such that for an
arbitrary property $a \in {\cal L}$:
\begin{equation} \label{joinstat}
\vee_jp_j \in \kappa(a) \Leftrightarrow p_j \in \kappa(a)\ \forall\ j
\end{equation}
If a state property space is property complete and state complete we
call it a `complete' state property space.
\end{definition}
The following definition and theorem explain why we have chosen to
call such a state property space complete.
\begin{definition} [complete pre-ordered set]
Suppose that $Z,\prec$ is a pre-ordered set. We say that
$Z$ is a complete pre-ordered set iff for each subset $(x_i)_i, x_i
\in Z$ of elements of $Z$ there exists an infimum and a
supremum in $Z$\footnote{An infimum of a subset
$(x_i)_i$ of a pre-ordered set $Z$ is an element of $Z$ that is smaller
than all the
$x_i$ and greater than any element that is smaller than all $x_i$. A
supremum of a subset $(x_i)_i$ of a pre-ordered set $Z$ is an element
of $Z$ that is greater than all the $x_i$ and smaller than any element
that is greater than all the $x_i$.}.
\end{definition}
\begin{theorem}
Consider a complete state property space $(\Sigma, {\cal L}, \xi,
\kappa)$. Then $\Sigma, \prec$ and ${\cal L}, \prec$ are complete
pre-ordered sets.
\end{theorem}
Proof: Consider an arbitrary set $(a_i)_i, a_i \in {\cal L}$. We
will show that $\wedge_ia_i$ is an infimum. First we have to proof that
$\wedge_ia_i \prec a_k\ \forall\ k$. This follows immediately from
(\ref{meetprop}) and the definition of $\prec$ given in
(\ref{ordprop}). Indeed, from this definition follows that we have to
prove that $\kappa(\wedge_ia_i) \subset \kappa(a_k)\ \forall\ k$.
Consider
$p \in \kappa(\wedge_ia_i)$. From (\ref{statprop}) follows that this
implies that $\wedge_ia_i \in \xi(p)$. Through (\ref{meetprop}) this
implies that $a_k \in \xi(p)\ \forall\ k$. If we apply 
(\ref{statprop}) again this proves that $p \in \kappa(a_k)\ \forall\
k$. So we have shown that $\kappa(\wedge_ia_i) \subset \kappa(a_k)\
\forall\ k$. This shows already that $\wedge_ia_i$ is a lower bound
for the set $(a_i)_i$. Let us now show that it is a greatest lower
bound. So consider another lower bound, a property $b \in
{\cal L}$ such that $b \prec a_k\ \forall\ k$. Let us show that $b 
\prec
\wedge_ia_i$. Consider $p \in \kappa(b)$, then we have $p \in
a_k\ \forall\ k$ since $b$ is a lower bound. This gives us that
$a_k \in \xi(p)\ \forall\ k$, and as a consequence $\wedge_ia_i \in
\xi(p)$. But this shows that $p \in \kappa(\wedge_ia_i)$. So we have
proven that $b \prec \wedge_ia_i$ and hence $\wedge_ia_i$ is an
infimum of the subset $(a_i)_i$. Let us now prove that $\vee_jp_j$ is
a supremum of the subset $(p_j)_j$. The proof is very similar, but we
use (\ref{joinstat}) in stead of (\ref{meetprop}). Let us again first
show that $\vee_jp_j$ is an upper bound of the subset $(p_j)_j$. We
have to show that $p_l \prec \vee_jp_j\ \forall\ l$. This
means that we have to prove that $\xi(\vee_jp_j) \subset \xi(p_l)\
\forall\ l$. Consider $a \in \xi(\vee_jp_j)$, then we have $\vee_jp_j
\in \kappa(a)$. From (\ref{joinstat}) it follows that $p_l \in
\kappa(a)\ \forall\ l$. As a consequence, and applying
(\ref{statprop}), we have that $a \in \xi(p_l)\ \forall\ l$. Let is now
prove that it is a least upper bound. Hence consider another upper
bound, meaning a state $q$, such that $p_l \prec q\ \forall\ l$. This
means that $\xi(q) \subset \xi(p_l)\ \forall\ l$. Consider now $a \in
\xi(q)$, then we have $a \in \xi(p_l)\ \forall\ l$. Using again
(\ref{statprop}), we have $p_l \in \kappa(a)\ \forall\ l$. From
(\ref{joinstat}) follows then that $\vee_jp_j \in \kappa(a)$ and
hence $a \in \xi(\vee_ja_j)$.

\par
 We have shown now that $\wedge_ia_i$ is
an infimum for the set $(a_i)_i, a_i \in {\cal L}$, and that
$\vee_jp_j$ is a supremum for the set $(p_j)_j, p_j \in \Sigma$.
It is a
mathematical consequence that for each subset $(a_i)_i, a_i \in {\cal
L}$, there exists also a supremum in ${\cal L}$, let is denote it by
$\vee_ia_i$, and that for each subset $(p_j)_j, p_j \in \Sigma$, there
exists also an infimum in
$\Sigma$, let us denote it by $\wedge_jp_j$. They are respectively
given by
$\vee_ia_i = \wedge_{x \in {\cal L}, a_i \prec x \forall i}\ x$ and
$\wedge_jp_j = \vee_{y \in \Sigma, y \prec p_j\forall j}\
y$\footnote{We remark that the supremum for elements of ${\cal L}$
and the infimum for elements of $\Sigma$, although they exists, as we
have proven here, have no simple operational meaning, as we will see
in the next section.}.

\par
\medskip
\noindent
For
both ${\cal L}$ and $\Sigma$ it can be shown that this implies that
there is at least one minimal and one maximal element. Indeed, an
infimum of all elements of ${\cal L}$ is a minimal element of ${\cal
L}$ and an infimum of the empty set is
a maximal element of ${\cal L}$. In an
analogous way a supremum of all elements of $\Sigma$ is a maximal
element of
$\Sigma$ and a supremum of the empty set
is a minimal element of $\Sigma$. Of
course there can be more minimal and maximal elements. If a property
$a \in {\cal L}$ is minimal we will express this by $a \approx 0$ and
if a property $b \in {\cal L}$ is maximal we will express this by $b
\approx I$. An analogous notation will be used for the maximal and
minimal states.

 For a
complete state property space we can specify the structure of the
maps $\xi$ and $\kappa$ somewhat more after having introduced the
concept of `property state' and `state property'.

\begin{theorem}
Consider a complete state property space $(\Sigma, {\cal L}, \xi,
\kappa)$. For $p \in \Sigma$ we define the `property state'
corresponding to $p$ as the property $s(p) = \wedge_{a \in
\xi(p)}a$. For $a \in {\cal L}$ we define the `state property'
corresponding to $a$ as the state $t(a) = \vee_{p \in \kappa(a)}p$.
We have two maps :
\begin{equation}
\begin{array}{ll}
t : {\cal L} \rightarrow \Sigma & a \mapsto t(a) \\
s : \Sigma \rightarrow {\cal L} & p \mapsto s(p)
\end{array}
\end{equation} 
and for $a, b \in {\cal L}$, and $(a_i)_i, a_i \in {\cal L}$ and
$p, q \in \Sigma$ and $(p_j)_j, p_j \in \Sigma$ we have :
\begin{equation}
\begin{array}{l}
a \prec b \Leftrightarrow t(a) \prec t(b) \\
p \prec q \Leftrightarrow s(p) \prec s(q) \\
t(\wedge_ia_i) \approx \wedge_it(a_i) \\
s(\vee_jp_j) \approx \vee_js(p_j)
\end{array}
\end{equation}
\end{theorem}
Proof: Suppose that $p \prec q$. Then we have $\xi(q) \subset
\xi(p)$. From this follows that $s(p) = \wedge_{a \in \xi(p)}a \prec
\wedge_{a \in \xi(q)}a = s(q)$. Suppose now that $s(p) \prec s(q)$.
Take $a \in \xi(q)$, then we have $s(q) \prec a$. Hence also $s(p)
\prec a$. But this implies that $a \in \xi(p)$. Hence this shows that
$\xi(q) \subset \xi(p)$ and as a consequence we have $p \prec q$.
Because $\wedge_ia_i \prec a_k\ \forall\ k$ we have $t(\wedge_ia_i)
\prec t(a_k)\ \forall k$. This shows that $t(\wedge_ia_i)$ is a lower
bound for the set $(t(a_i))_i$. Let us show that it is a smallest
lower bound.
Suppose that $p \prec t(a_k)\ \forall\ k$. We remark that $t(a_k)
\in \kappa(a_k)$. Then it follows that $p \in \kappa(a_k)\ \forall\
k$. As a consequence we have $a_k \in \xi(p)\ \forall\ k$. But then
$\wedge_ia_i \in \xi(p)$ which shows that $p \in
\kappa(\wedge_ia_i)$. This proves that $p \prec t(\wedge_ia_i)$. So
we have shown that $t(\wedge_ia_i)$ is a smallest lower bound and
hence it is equivalent to $\wedge_it(a_i)$.

\begin{theorem} \label{interval}
Consider a complete state property space $(\Sigma, {\cal L}, \xi,
\kappa)$. For $p \in \Sigma$ we have $\xi(p)
= [s(p), +\infty] = \{ a \in {\cal L} \
\vert \ s(p) \prec a\}$. For
$a
\in {\cal L}$ we have $\kappa(a) = [-\infty,
t(a)] = \{p \in \Sigma\ \vert\ p \prec t(a)\}$.
\end{theorem}
Proof: Consider
$b \in [s(p), +\infty]$. This means that $s(p) \prec b$, and hence $b
\in
\xi(p)$. Consider now $b
\in \xi(p)$. Then $s(p) \prec b$ and hence $b \in [s(p), +\infty]$.

\par
\medskip
\noindent
If $p$ is a state such that $\xi(p) = \emptyset$, this means that
there is no property actual for the entity being in state $p$. We
will call such states `improper' states. Hence a `proper' state is a
state that makes at least one property actual. In an analogous way,
if $\kappa(a) = \emptyset$, this means that there is no state that
makes the property $a$ actual. Such a property will be called an
`improper' property. A `proper' property is a property that is actual
for at least one state.

\begin{definition}
Consider a state property space $(\Sigma, {\cal L}, \xi,
\kappa)$. We call $p \in \Sigma$ a `proper' state iff $\xi(p) \not=
\emptyset$. We call $a \in {\cal L}$ a `proper' property iff
$\kappa(a) \not= \emptyset$. A state $p \in \Sigma$ such that $\xi(p)
= \emptyset$ is called an `improper' state, and a property $a \in {\cal
L}$ such that $\kappa(a) = \emptyset$ is called an `improper'
property. 
\end{definition}
It easily follows from theorem \ref{interval} that a complete state
property space has no improper states ($I \approx \wedge\emptyset \in
\xi(p)$) and no improper properties ($0 \approx \vee \emptyset \in
\kappa(a)$).

\subsection{Tests and preparations: the
operational part}

\noindent
Our contact with physical entities of the exterior world happens by
means of experiments we can perform. A test is an
experiment we perform on the physical entity in a certain
state testing a certain hypothesis. States can often be
prepared. A preparation is an experiment we perform on the
physical entity such that as a result of the experiment the entity is
in a certain state.
We will not develop the algebra of experiments connected to a
physical entity in a complete way in this paper, and refer to [6] for
such an elaboration. Here we will only introduce the concepts that
we need for our principal purpose: the presentation of quantum
axiomatics. 

\par
\medskip
\noindent {\it i) The tests on the entity $S$}

\par
\medskip
\noindent
Tests are experiments that verify a certain hypothesis about the
entity $S$. More specifically tests can test properties of the
entity $S$ in the following way. With a property $a \in
{\cal L}$ corresponds a test $\alpha(a)$, which is an experiment
with two possible outcomes `yes' and `no'. If the test $\alpha(a)$
has an outcome `yes' it does not yet prove that the property
$a$ is actual. It is only when we can predict
 with certainty that the test would have an outcome `yes',
without necessarily performing it, that the property
$a$ is actual.

\begin{definition} [testing a property]
Suppose that we have an entity $S$ with corresponding state property
space $(\Sigma, {\cal L}, \xi, \kappa)$. $\alpha(a)$ is a test of
the property $a \in {\cal L}$ if we have
\begin{eqnarray}
a \in \xi(p) &\Leftrightarrow& {\rm `yes'\ can\ be\
predicted\ for\ } \alpha(a) \\ \nonumber
& & S\ {\rm being\ in\ state\ } p
\end{eqnarray}
\end{definition} 

\par
\medskip
\noindent {\it ii) The preparations of the entity $S$}

\par
\medskip
\noindent
Preparations are experiments that prepare a state of the entity $S$.
More specifically, with a state $p \in \Sigma$ corresponds a
preparation $\pi(p)$ which is an experiment such that after the
performance of the experiment the entity `is' in state $p$.

\begin{definition} [preparing a state]
Suppose that we have an entity $S$ with corresponding state property
space $(\Sigma, {\cal L}, \xi, \kappa)$. $\pi(p)$ is a preparation of
the state
$p
\in
\Sigma$ if we have
\begin{equation}
p \in \kappa(a) \Leftrightarrow a\ {\rm is\ actual\ after\ the\
preparation\ } \pi(p)
\end{equation}
\end{definition}
For a set of tests $(\alpha_i)_i$ and for a set of preparations
$(\pi_j)_j$ we can now introduce in a very natural way a new test,
that we call the product test, denoted by $\Pi_i\alpha_i$, and a new
preparation, that we call the product preparation, denoted by
$\Pi_j\pi_j$, as follows:

\begin{definition} [product test and preparation]
\label{productteststate} To execute $\Pi_i\alpha_i$ we choose one of
the
$\alpha_i$, perform it and consider the outcome that we obtain. To
execute
$\Pi_j\pi_j$ we choose one of the $\pi_j$, perform is and consider
the state that we obtain.
\end{definition}
We want to show now that the product test tests an infimum of a set
of properties, while the product preparation prepares a supremum of
a set of states.

\begin{theorem}
Suppose that we have an entity $S$ with corresponding state property
space $(\Sigma, {\cal L}, \xi, \kappa)$. Consider a set of properties
$(a_i)_i, a_i \in {\cal L}$ and a set of states $(p_j)_j, p_j \in
\Sigma$. Suppose that we have tests and preparations available for all
properties and states. Then the product test $\Pi_i\alpha(a_i)$ tests
a meet property $\wedge_ia_i$, where $\alpha(a_k)$ tests $a_k\
\forall\ k$, and the product preparation
$\Pi_j\pi(p_j)$ prepares a join state $\vee_jp_j$, where $\Pi(p_l)$
prepares $p_l\ \forall\ l$.
\end{theorem}
Proof: We have to show that `yes can be predicted for
$\Pi_i\alpha(a_i)$ the entity $S$ being in state $p$' is equivalent to
`$a_k \in \xi(p)\ \forall\ k$'. This follows immediately from the
definition of the product test. Indeed `yes can be predicted
for $\Pi_i\alpha(a_i)$ the entity $S$ being in state $p$' is equivalent
to `yes can be predicted for $\alpha(a_k)\ \forall\ k$ the entity $S$
being in state $p$'. Consider now an arbitrary property $a \in {\cal
L}$ and suppose that $(\pi(p_j))_j$ is a set of preparations that make
$a$ actual if the entity $S$ is in state $p_j$. Consider now the
preparation $\Pi_j\pi(p_j)$, that consists of choosing one of the
$\pi(p_j)$ and performing it. Then it is  clear that $a$ is actual
after this preparation, since $S$ will be in one of the states $p_j$.
On the other hand, suppose now that $\Pi_j\pi(p_j)$ is a preparation
that makes $a \in {\cal L}$ actual. Consider an arbitrary one of the
preparations $\pi(p_k)$ of the product preparation. Then obviously
also this preparation has to make $a$ actual, since it could have
been this one that was chosen by performing the product preparation.
This shows that $\Pi_j\pi(p_j)$ prepares the state $\vee_jp_j$.

\par
\medskip
\noindent
This theorem shows that it is natural to introduce the meet property
for a set of properties and the join state for a set of states, like
we did in the foregoing section.
It is time now that we expose the concepts that we have introduced
here for the example of the spin of a spin${1 \over 2}$ quantum
particle.

\subsection{The example of the spin model}
\label{complquant}

\noindent
In section \ref{quantspin} we have explained in detail the standard
quantum description of the spin of a spin${1 \over 2}$ quantum
particle. For the case of the spin of a spin${1 \over 2}$
quantum particle, the experiments $f_u$ have only two outcomes
$o_1$ and $o_2$, and hence they are tests in the sense that if $o_1$
is interpreted as `yes' then $o_2$ is `no'. This means that we can
represent the properties by means of the projection operators that we
use to represent the experiments. For each direction
$u$ we have a property $a_u$ that is represented by the projection
operator given in formula (\ref{spinproj01})
\begin{equation} 
P_u = {1 \over 2} \left(
\begin{array}{ll}
1 + \cos\alpha & e^{-i\beta}\sin\alpha \\
e^{+i\beta}\sin\alpha & 1 - \cos\alpha
\end{array} \right)
\end{equation}
where $u = (1,\alpha,\beta)$ and hence $\alpha$ and $\beta$ are the
spherical coordinates angles of the vector $u$. The set of properties
${\cal L}_{spin{1
\over 2}}$ is given by these properties $a_u$ and the maximal
and minimal property that we will respectively denote by $I$ and
$0$\footnote{It will become clear in the next section why for the
quantum case there is a unique minimal property and a unique maximal
property}. Hence:
\begin{equation} \label{propspin}
{\cal L}_{spin{1 \over 2}} = \{a_u, I, 0\ \vert\ u \in\ {\rm surface\
of\ the\ sphere}\ ball \} 
\end{equation}
We have stated in section
\ref{quantspin} that a state $q_v$ of the spin of a spin${1 \over 2}$
quantum particle in direction $v$ is represented by means of the unit
vector $c_v$ in the two dimensional complex vector space $\Bbb{C}^2$. We 
have to elaborate a little bit more on the description of
the states. Indeed, if we consider again our quantum machine, which
is a model for the spin of a spin${1 \over 2}$ quantum particle, then
we can see that there are more states than the ones represented by
the unit vectors. If we consider a point $w$ in the interior of the
sphere, hence not on the surface of the sphere, then this is also a
possible state of the quantum machine, not represented however by
a unit vector of the vector space, but by a density
operator\footnote{A density operator is an operator $W$
such that $<c, Wc> = <W c,c>\ \forall\ c \in \Bbb{C}^2$ (which
means self-adjointness), and such that $0 \le \ <Wc,Wc>\ \forall\ c
\in \Bbb{C}^2$ (which means positiveness), and such that the
trace equals to 1.}. Let us analyse this situation in detail.

Let us calculate the probabilities for such a state $p_w$ where
$w$ is a point inside the sphere. First we remark the
following. Because $ball$ is a convex set, each vector
$w \in ball$ can be written as a convex linear combination of two
vectors $v$ and $-v$ on the surface of the sphere (see Fig 13). More
concretely this means that we can write (referring to the $w$ and $v$
and $-v$ in the Figure 13):
\begin{equation}
w = \lambda_1 \cdot v - \lambda_2 \cdot v,\ \  0 \leq \lambda_1,
\lambda_2
\leq 1,
\
\  \lambda_1 + \lambda_2 = 1
\end{equation} 
Hence, if we introduce these convex combination coefficients
$\lambda_1, \lambda_2$  we have $w = (\lambda_1-\lambda_2) \cdot v$.
Let us calculate now the transition probabilities in a general state
$p_w$ with $w
\in ball$ and hence
$\|w\| \leq 1$ (see Fig 13). Again the probability
$\mu(e_u, p_w, o_1)$, that the particle $P$ ends up in point $u$ 
and hence experiment $e_u$ gives outcome
$o_1$ is given by the length of the piece of elastic
$L_1$ divided by the total length of the elastic. The probability, 
$\mu(e_u, p_w, o_2)$, that the particle $P$ ends up in point 
$-u$, and hence experiment $e_u$ gives outcome $o_2$ is
given by  the length of the piece of elastic $L_2$ divided by the
total length of the elastic. This means that we have:
\begin{eqnarray} \label{probform}
\mu(e_u, p_w, o_1) &=& {L_1\over 2} = {1 \over 2}(1 +
(\lambda_1-\lambda_2)  \cos\theta) \\
 &=& \lambda_1 \cos^2{\theta\over 2}
+ \lambda_2 \sin^2{\theta\over 2} \\
\mu(e_u, p_w, o_2) &=& {L_2\over 2} = {1 \over 2}(1 -
(\lambda_1-\lambda_2)  \cos\theta) \\
 &=& \lambda_1 \sin^2{\theta\over 2}
+
\lambda_2 \cos^2{\theta\over 2}
\end{eqnarray}

\vskip 0.5 cm

\hskip 3.7 cm \includegraphics{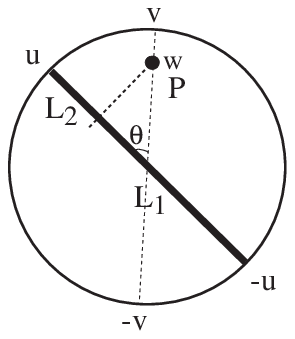}

\begin{quotation}
\noindent \baselineskip= 7pt \smallroman Fig 13 : A representation of the 
experimental process in the case of a state $\scriptstyle p_w$ where
$\scriptstyle w$ is a point of the interior of the sphere. The elastic
of length 2, corresponding to the experiment
$\scriptstyle e_u$, is installed between $\scriptstyle u$ and 
$\scriptstyle -u$. The probability, $\scriptstyle \mu(e_u, p_w,
o_1)$, that the particle  $\scriptstyle P$ ends up in point
$\scriptstyle u$ under influence of the experiment
$\scriptstyle e_u$ is given by the length of the piece of
elastic $\scriptstyle L_1$ divided by the total length of the elastic.
The probability,
$\scriptstyle
\mu(e_u, p_w, o_2)$, that the particle $\scriptstyle P$ ends up in
point $\scriptstyle -u$ is given by the length of the piece of elastic
$\scriptstyle L_2$ divided by the total length of the elastic.
\end{quotation}

\medskip
\noindent
These are new probabilities that will never be obtained if we limit
the set of states to the unit vectors of the two dimensional complex
space. The question is now the following: can we find a mathematical
concept, connected in some way or another to the Hilbert space, that
would allow us, with a new quantum rule for calculating probabilities,
to recover these probabilities? The answer is yes. We will show that
these new `pure' states of the interior of the sphere can be
represented using density operators, the same operators that are used
within standard quantum formalism to represent mixed states. And the
standard quantum mechanical formula that is used to calculate
probabilities connected to mixed states, represented by density
matrices, can also be used to calculate the probabilities that we have
identified here. But of course the meaning will be different: in our
case this standard formula will represent a transition probability
from one pure state to another and not the probability connected to
the change of a mixed state. Let us show all this explicitly and do
this by constructing the density matrices in question.

The well known quantum formula for the calculation of
transition probabilities  related to an experiment
$e$, represented by the projections $\{P, I - P\}$, and where
the quantum entity is in a mixed state
$p$ represented by the density operator
$W$, is the following:
\begin{equation}
\mu(e, p, P) = tr(W \cdot P) \label{probcal}
\end{equation}
where $tr$ is the trace of the operator\footnote{The trace of an
operator is the sum of its diagonal elements.}.

A standard quantum mechanical calculation shows that the density 
operator representing the ray state $c_v = (\cos{\theta
\over 2}  e^{i{\phi
\over 2}}, \sin{\theta \over 2}  e^{-i{\phi \over 2}})$ (see 
(\ref{raystate})) is given by:
\begin{equation}
W(v)  = \left( \begin{array}{ccc} 
\cos^2{\theta \over 2} & \sin{\theta \over 2}\cos{\theta \over 2}e^{-
i\phi}  \\
\sin{\theta \over 2}\cos{\theta \over 2}e^{i\phi} & \sin^2{\theta \over 
2} 
\end{array} \right)
\end{equation}
and the density operator representing the diametrically opposed ray 
state $c_{-v}$ is given by:
\begin{equation}
W(-v)  = \left( \begin{array}{ccc} 
\sin^2{\theta \over 2}  & -\sin{\theta \over 2}\cos{\theta \over
2}e^{-i\phi}  \\ -\sin{\theta \over 2}\cos{\theta \over 2}e^{i\phi} &
\cos^2{\theta \over 2} 
\end{array} \right)
\end{equation}
We will show now that the convex linear combination of these two 
density operators with convex weights $\lambda_1$ and
$\lambda_2$ represents the state $p_w$ if we use the standard quantum 
mechanical formula (\ref{probcal}) to calculate the transition
probabilities. If, for $w = \lambda_1v + \lambda_2(-v)$, we put :
\begin{equation}
W(w) = \lambda_1W(v) + \lambda_2 W(-v)
\end{equation}
we have:
\small
\begin{equation} \label{densop}
W(w) = \left( \begin{array}{ccc} 
\lambda_1\cos^2{\theta \over 2} + \lambda_2\sin^2{\theta \over 2} & 
(\lambda_1-\lambda_2)\sin{\theta
\over 2}\cos{\theta \over 2}e^{-i\phi}  \\
(\lambda_1-\lambda_2)\sin{\theta
\over 2}\cos{\theta \over 2}e^{i\phi} & \lambda_1\sin^2{\theta \over 2} +
\lambda_2\cos^2{\theta
\over 2} 
\end{array} \right)
\end{equation}
\normalsize
and it is easy to calculate now the transition probabilities 
using (\ref{probcal}) and:
\begin{equation}
P = \left( \begin{array}{cc}
1 & 0 \\
0 & 0
\end{array}
\right)
\end{equation}
We have:
\begin{equation}
W(w) \cdot P = \left( \begin{array}{ccc} 
\lambda_1\cos^2{\theta \over 2} + \lambda_2\sin^2{\theta \over 2} & 0  \\
(\lambda_1-\lambda_2)\sin{\theta \over 2}\cos{\theta \over 2}e^{i\phi} &
0 
\end{array} \right)
\end{equation}
and hence, comparing with (\ref{probform}), we find:
\begin{equation}
tr(W(w) \cdot P) = \lambda_1\cos^2{\theta \over 2} +
\lambda_2\sin^2{\theta 
\over 2} =
\mu(e_u, p_w, o_1)
\end{equation}
In an analogous way we find that:
\begin{eqnarray}
tr(W(w) \cdot (I-P)) &=& \lambda_1\sin^2{\theta \over 2} +
\lambda_2\cos^2{\theta 
\over 2} \\ \nonumber
& & =
\mu(e_u, p_w, o_2)
\end{eqnarray}
So we have shown that we can represent each one of the new states
$p_w$ by the density operator $W(w)$ if we use (\ref{probcal})
for the calculation of the transition probabilities.

We can also prove that each density operator in $\Bbb{C}^2$ is
of this form. We show this easily by using the
general properties of density operators.

Let us identify now the set of states for the
case of the spin of a spin${1 \over 2}$ quantum particle. Each state
is of the form $p_w$ with $w$ a point of the sphere $ball$. There also
exist a zero state $0$ and a unit state $I$. 

\begin{equation} \label{statspin}
\Sigma_{spin{1 \over 2}} = \{p_w, 0, I\ \vert\ w \in ball\} 
\end{equation}

The state property space corresponding to the spin of a spin${1 \over
2}$ quantum particle is given by $(\Sigma_{spin{1 \over 2}}, {\cal
L}_{spin{1 \over2}}, \xi_{spin{1 \over 2}}, \kappa_{spin{1 \over
2}})$, where
\begin{equation}
\xi_{spin{1 \over 2}}: \Sigma_{spin{1 \over 2}} \rightarrow {\cal
P}({\cal L}_{spin{1 \over 2}}) 
\end{equation}
We have for $u$ and $v$ belonging to the sphere $ball$:
\begin{equation} \label{spinstat01}
a_u \in \xi_{spin{1 \over 2}}(p_w) \Leftrightarrow p_w
\in \kappa_{spin{1 \over 2}}(a_u) \Leftrightarrow u = w 
\end{equation}
We have, for $|v| =1$, and hence pertaining to the surface of $ball$:
\begin{equation} \label{spinxi01}
\xi_{spin{1 \over 2}}(p_v) = \{a_v, I\}
\end{equation}
For $w < 1$, and hence pertaining to the interior of $ball$ we
have:
\begin{equation} \label{spinxi02}
\xi_{spin{1 \over 2}}(p_w) = \{I\}
\end{equation}

\section{Quantum axiomatics: the technical part}

\noindent
In the foregoing section we have introduced the structure that can be
operationally founded. To come to the full structure of standard
quantum mechanics some additional axioms have to be introduced which
are more of a technical nature. This section will be mathematically
more sophisticated, but can be skipped
for those readers that are mainly interested in the results that are
presented in next section.
\par
In the introduction we mentioned that quantum
axiomatics was developed to build up standard quantum mechanics and
not to change it. Meanwhile it has become clear that some
of the more technical axioms of standard quantum mechanics are probably
not generally satisfied in nature. This finding will be our main
comment regarding the standard axioms, and for this reason we anyhow
have to introduce them. An analysis of the failing axioms and their
consequences is presented in the next section.

\subsection{State property systems}

\noindent
Since we will introduce in this section the axioms that have very
little operational meaning, we will enter much less in detail.

\par
\bigskip
\noindent
{\it (1) The identification of properties}

\par
\medskip
\noindent
As we have seen in the example of the spin of a spin${1 \over 2}$
quantum particle, a property of the physical entity is represented by
a projection operator. This remains true for a general quantum
entity. Suppose that we consider two properties $a$ and $b$
of a general quantum entity, and their corresponding projection
operators
$P_a$ and $P_b$. The implication of properties, giving rise to a
pre-order relation on ${\cal L}$, is translated for a quantum entity
as follows by means of the projection operators:
\begin{theorem} \label{quant01}
Consider a state property space $(\Sigma, {\cal L},
\xi, \kappa)$ corresponding to a quantum entity $S$, described by
means of the standard quantum formalism in a Hilbert space ${\cal
H}$. For two properties $a, b \in {\cal L}$, and corresponding
projection operators $P_a$ and $P_b$ we have:
\begin{equation}
a \prec b \Leftrightarrow P_a P_b = P_b P_a = P_a
\end{equation}
\end{theorem}
\begin{definition}
Consider a pre-ordered set $Z, \prec$. The pre-order relation $\prec$
is a partial order relation iff we have for $x, y \in Z$
\begin{equation}
x \prec y\ {\rm and}\ y \prec x\ \Rightarrow\ x = y
\end{equation}
\end{definition}
\begin{theorem} \label{propid}
Consider a state property space $(\Sigma, {\cal L},
\xi, \kappa)$ corresponding to a quantum entity $S$, described by
means of the standard quantum formalism in a Hilbert space ${\cal
H}$. For two properties $a, b \in {\cal L}$ we have:
\begin{equation}
a \prec b\ \ {\rm and} \ \ b \prec a \ \Rightarrow\ a = b
\end{equation}
and hence the pre-order relation on ${\cal L}$ is a partial
order relation.
\end{theorem}
Proof: Suppose that $a \prec b$ and $b \prec a$. Then from
theorem~\ref{quant01} follows that $P_a = P_a P_b = P_b P_a = P_b$.
which means that $a =b$.

\par
\bigskip
\noindent
{\it (2) Completeness of the property lattice}

\par
\medskip
\noindent
The second special property for quantum entities that we will
identify has already been explained in the foregoing section. It is
related to the existence of the meet property for a set of properties.
Suppose that we consider again a quantum entity and a set of properties
$(a_i)_i$ with corresponding set of projection operators
$(P_{a_i})_i$. Then there exists a unique projection operator
$P_a$ that corresponds to the meet property $\wedge_ia_i$.
\begin{theorem} \label{complet}
Consider a state property space $(\Sigma, {\cal L},
\xi, \kappa)$ corresponding to a quantum entity $S$, described by
means of the standard quantum formalism in a Hilbert space ${\cal
H}$. For a set of properties $(a_i)_i, a_i \in {\cal L}$, and
corresponding set of projection operators $(P_{a_i})_i$, there exists
a projection operator $P_a$ such that
\begin{equation}
a \in \xi(p) \Leftrightarrow a_i \in \xi(p)\ \forall\ i
\end{equation}
which means that $a = \wedge_ia_i$ and which means that the state
property space is property complete (see
definition
\ref{comstatprop})
\end{theorem}
We will not prove this theorem because it will lead us into
too many technical details. We only mention that this
is a well known result about Hilbert space projectors. 

\par
\bigskip
\noindent
{\it (3) Minimal property and maximal property}

\par
\medskip
\noindent
We can remark now also that for the case of a quantum entity the
maximal property $I$ is always actual for any state of the entity and
the minimal property $0$ is never actual (both are unique since the
set of properties is a partially ordered set). Indeed:

\begin{theorem}
Consider a state property space $(\Sigma, {\cal L},
\xi, \kappa)$ corresponding to a quantum entity $S$, described by
means of the standard quantum formalism in a Hilbert space ${\cal
H}$. Consider an arbitrary state $p_W \in \Sigma$ and let $0$ be the
minimal property and $I$ be the maximal property of ${\cal L}$, then
$0 \not\in \xi(p)$ and $I \in \xi(p_W)$.
\end{theorem}
Proof: The minimal property is represented in quantum mechanics
by the zero projection $0$, and the maximal property by the unit
operator $I$. We have $tr(W \cdot 0) = 0$ and $tr(W \cdot I) = tr(W) =
1$ which shows that $0 \not\in \xi(p_W)$ and $I \in \xi(p_W)$.

\par
\medskip
\noindent
The additional structure that we have observed for a quantum entity in
(1), (2) and (3) will be our inspiration for the first axiom, and
will make us introduce the structure of a state property system, that
we have studied already intensively [30] and [31].

\begin{definition} [state property system]
Suppose that we have a state property space $(\Sigma, {\cal L}, \xi,
\kappa)$. This state property space is a state property system iff
$({\cal L},
\prec,
\wedge,
\vee)$ is a complete lattice\footnote{A complete lattice is a complete
partially ordered set.}, and for
$p
\in
\Sigma$,
$I$ the maximal element and
$0$ the minimal element of ${\cal L}$, and $a_i \in {\cal L}$, we have:
\begin{equation}
\begin{array}{lll}
I \in \xi(p) &
O \not\in \xi(p) &
a_i \in \xi(p)\ \forall\ i \Leftrightarrow \wedge_ia_i \in \xi(p)
\end{array}
\end{equation}
\end{definition}
\begin{theorem}
Consider a state property space $(\Sigma, {\cal L},
\xi, \kappa)$ corresponding to a quantum entity $S$, described by
means of the standard quantum formalism in a Hilbert space ${\cal
H}$, then $(\Sigma, {\cal L},
\xi, \kappa)$ is a state property system.
\end{theorem}

The foregoing results make it possible to introduce the first
axiom:
\begin{axiom} [state property system]
Suppose that we have a state property space $(\Sigma, {\cal L},
\xi, \kappa)$ corresponding to an entity $S$. We say that axiom 1 is
satisfied iff the state property space is a state property system.
\end{axiom}
If axiom 1 is satisfied we will call the set of
properties the `property lattice' corresponding to entity $S$.

\subsection{Atomic states} \label{linear}

\noindent
We have seen that the set of states $\Sigma$ of a general physical
entity has a natural pre-order relation that we have called the state
implication. We have also explained that a state in standard quantum
mechanics can be represented by a density operator. Some of the density
operators represent vector states of the Hilbert space. The
representation theorem that we will put forward in this section
is inspired on the classical representation theorem formulated for
the situation where all states are vector states. Therefore we wonder
whether it is possible to characterize the vector states in a more
general way. This is indeed possible by means of the mathematical
concept of `atom' that we will introduce now. 
\begin{definition}
Consider a pre-ordered set $Z, \prec$. We say
that $x \in Z$ is an `atom' iff $x$ is not a
minimal element and for 
$y \prec x$ we have
$y \approx x$ or $y$ is a minimal element of $Z$.
\end{definition}
\begin{definition}
Consider a state property space $(\Sigma, {\cal L}, \xi, \kappa)$
describing an entity $S$. We know that $\Sigma, \prec$ is a
pre-ordered set. We call
$p \in \Sigma$ an `atomic state' of the entity $S$ iff $p$ is an atom
for the pre-order relation on $\Sigma$. We will denote the set
of atomic states of a state property space by means of $\Lambda$.
\end{definition}
Let us first investigate which are the atomic states for our example
of the quantum machine. Following (\ref{spinxi01}) and
(\ref{spinxi02}) we have $p_v \prec p_w$ and $p_v \not\approx p_w$ for
$|v| = 1$ and
$w < 1$. This shows that none of the states $p_w$ with $|w| < 1$ is an
atomic state. From (\ref{spinstat01}) follows that $p_v \prec p_w$ for
$|v| = |w| = 1$ implies that $v = w$ and hence $p_v = p_w$. This shows
that all of the states $p_v$ with $|v| = 1$ are atomic states.
Hence the atomic states for the quantum machine are exactly these
states that correspond to points on the surface of $ball$.
For the situation of a general quantum entity described in a Hilbert
space ${\cal H}$ a similar result can be shown: the atomic states
are those states that are represented by density operators that
correspond to vectors of the Hilbert space.
\begin{theorem} \label{densatom}
Consider a state property space $(\Sigma, {\cal L}, \xi, \kappa)$
corresponding to a quantum entity $S$ described by a Hilbert space
${\cal H}$. A state $p \in \Sigma$ is atomic iff it is represented by a
density operator corresponding to a vector of the Hilbert space.
\end{theorem}
Proof: Consider a state $p_W \in \Sigma$ represented by a density
operator $W$ of ${\cal H}$. This density operator $W$ can always be
written as a convex linear combination of density operators $W_i$
corresponding to vectors of ${\cal H}$ and representations of states
$p_{W_i} \in \Sigma$\footnote{This is a well known property of density
operators in a Hilbert space.}. Hence we have
$W =
\sum_i\lambda_iW_i$. Consider a property $a \in {\cal L}$ such that
$a \in \xi(p_W)$. This means that $tr(WP_a) = 1$ where $P_a$ is
the projection operator representing the property $a$. We have
$tr(WP_a) = tr(\sum_i\lambda_iW_iP_a) = \sum_i\lambda_itr(W_iP_a)$.
Since $0 \le \lambda_i \le 1\ \forall\ i$ and $\sum_i\lambda_i = 1$ and
$0
\le tr(W_iP_a)
\le 1\ \forall\ i$ this proves that
$tr(W_iP_a) = 1\ \forall\ i$. As a consequence we have $a \in
\xi(p_{W_i})$. This shows that $\xi(p_W) \subset \xi(p_{W_i})$ and
hence $p_{W_i} \prec p_W$. This shows already that genuine density
operators that are not corresponding to a vector of the Hilbert space
are not atomic. Suppose now that we consider a state $p_W \in \Sigma$
where $W$ corresponds to a vector $c \in {\cal H}$ and another state
$p_V \in \Sigma$ such that $p_V \prec p_W$. This means that $\xi(p_W)
\subset \xi(p_V)$. Suppose the property represented by the projector
operator on the vector
$c$, let us denote it $P_c$, is contained in $\xi(p_W)$ and hence also
in
$\xi(P_V)$. From this follows that $V = P_c$ and hence $p_V = p_W$.
This shows that states represented by density operators
corresponding to vectors of the Hilbert space are atomic states.

\par
\medskip
\noindent
The reader has perhaps meanwhile understood in which way we will
gradually arrive at a full axiomatization of standard quantum
mechanics. We analyse step by step what are the requirements that are
additionally satisfied for the state property space of a quantum
entity described by standard quantum mechanics. To proceed along this
line we will now first show that for an entity satisfying axiom
1,  the atomic states can be identified unequivocally with
atomic properties of the property lattice. This will make it
possible to concentrate only on the structure of the property lattice.

\begin{theorem}
Consider a state property space
$(\Sigma, {\cal L}, \xi, \kappa)$ satisfying axiom 1 (hence a state
property system) with set of atomic states $\Lambda$. Let us denote
by ${\cal A}$ the set of atoms of ${\cal L}$. If we consider the
function $s : \Sigma \rightarrow {\cal L}$, then $s(\Lambda) = {\cal
A}$.
\end{theorem}
Proof: Suppose that $r \in \Lambda$ and let is show that $s(r)$ is a
atom of ${\cal L}$. Consider $a \in {\cal L}$ such that $a \prec
s(r)$ and $a \not= 0$. We remark that in this case $\kappa(a) \not=
\emptyset$. Indeed, suppose that $\kappa(a) = \emptyset$ then $a
\prec b\ \forall\ b \in {\cal L}$ and hence $a = 0$. If $\kappa(a)
\not= \emptyset$ there exists a $p \in \Sigma$ such that $p \in
\kappa(a)$. We then have $a \in \xi(p)$ and as a consequence $s(p)
\prec a \prec s(r)$. From this follows that $p \prec r$, but since $r
\in \Lambda$ we have $p \approx r$. This implies that $s(p) = s(r)$
and hence $a = s(r)$. This proves that $s(r)$ is an atom of ${\cal L}$.
Consider now $a \in {\cal A}$. Let us show that there exists a $p \in
\Lambda$ such that $s(p) = a$. We have $\kappa(a) \not= \emptyset$
since $a \not= 0$. This means that there exists $p \in \Sigma$ such
that $p \in \kappa(a)$. Hence $a \in \xi(p)$ and as a consequence we
have $s(p) \prec a$. From this follows that $s(p) = 0$ or $s(p) = a$.
We remark that $s(p) = 0$ is not possible since this would imply that
$\xi(p)  = {\cal L}$ and hence $0 \in \xi(p)$ which is forbidden.
Hence we have $s(p) = a$. We must still show that $p$ is an atom.
Indeed suppose that $r \prec p$, then $s(r) \prec s(p) = a$. This
shows that $s(r) = a$ and hence $\xi(r) = \xi(p)$.

\par
\medskip
\noindent
The foregoing theorem shows that we can represent an atomic state
by means of the atom of the property lattice on which it is mapped
by the map $s$. That is the reason we will concentrate on the
property lattice ${\cal L}$. It could well be possible that
no or very few atomic states exist. For the case of standard quantum
mechanics there are however many atomic states.
\begin{theorem}
Consider a state property space $(\Sigma, {\cal L}, \xi, \kappa)$
describing a standard quantum mechanical entity $S$ in a Hilbert
space ${\cal H}$. Each property $a_P \not= 0$ represented by a
non zero projection operator $P$ equals the supremum of the set of
rays (one dimensional projections) contained in $P$.
\end{theorem}
This is the inspiration for the next axiom. First we introduce the
concept of a complete atomistic lattice.
\begin{definition}
Consider a complete lattice ${\cal L}, \prec$ and its set of atoms
${\cal A}$. We say that
${\cal L}$ is `atomistic' iff each $a \in {\cal L}$ is equal to the
supremum of its atoms, i.e. $a = \vee_{c \in {\cal A}, c \prec a}c$.
\end{definition}
\begin{axiom} [atomicity]
Consider a state property system $(\Sigma, {\cal L}, \xi, \kappa)$
describing an entity $S$. We say that the state property system
satisfies axiom 2 iff its property lattice is atomistic.
\end{axiom}
We know from the foregoing that the atoms of ${\cal L}$ represent the
atomic states of the entity $S$.

Apart from atomicity, the property lattice of an entity described 
by standard quantum mechanics satisfies an additional property,
called the `covering law'. It is the following:

\begin{axiom} [covering law]
The property lattice ${\cal L}$ of a state property system $(\Sigma,
{\cal L}, \xi, \kappa)$ satisfies the `covering law' iff for $c \in
{\cal A}$ and $a, b \in {\cal L}$ such that $a \wedge c = 0$ and $a
\prec b \prec a \vee c$ we have $a = b$ or $a \vee c = b$.
\end{axiom}

The covering law demands that the supremum of a property and an atom
`covers' this property, in the sense that their does not exists a
property in between.

\par
The first three axioms introduce the linearity of
the set of states of the entity. Indeed it can be shown that a
complete atomic lattice satisfying the covering law and containing
sufficiently many atoms is isomorphic to a projective geometry. Making
use of the fundamental theorem of projective geometry we can construct
of vector space coordinating this geometry and also representing the
original lattice of properties. We refer to [34] for a proof of this
fundamental representation theorem for complete atomic lattices
satisfying the covering law.

\subsection{Orthogonality}

\noindent
The next axiom is inspired by the specific and strong orthogonality
structure that exists on a Hilbert space. If axiom 1 is
satisfied the set of properties is a complete lattice. We give now the
definition of the structure of an orthocomplementation, which will
make it possible for us to introduce the next axiom.

\begin{definition} [orthocomplementation]
Consider a complete lattice ${\cal L}$. We say that $':{\cal L}
\mapsto {\cal L}$ is an orthocomplementation iff for $a, b \in {\cal
L}$ we have:
\begin{equation}
\begin{array}{l}
a \prec b \Rightarrow b' \prec a' \\
(a')' = a \\
a \wedge a' = 0
\end{array}
\end{equation}
\end{definition}
\begin{theorem}
Consider a state property space $(\Sigma, {\cal L},
\xi, \kappa)$ corresponding to a quantum entity $S$, described by
means of the standard quantum formalism in a Hilbert space ${\cal
H}$. If we define for a property $a \in {\cal L}$, and corresponding
projection operator $P_a$, the property $a'$ as corresponding to the
projection operator $I - P_a$, then $': {\cal L} \mapsto {\cal L}$ is
an orthocomplementation.
\end{theorem}
Proof: Suppose that we have $a, b \in {\cal L}$ and their corresponding
projection operators $P_a$ and $P_b$. If $a \prec b$ then $P_a = P_a
P_b = P_b P_a$. We have $(I-P_a)(I-P_b) = I - P_b - P_a + P_a P_b = I
- P_b = I - P_b - P_a + P_bP_a = (I-P_b)(I-P_a)$. This shows that $b'
\prec a'$. We have $I - (I - P_a) = P_a$ which proves that $(a')'
=a$. Since $P_a(I-P_a) = 0$ we have $a \wedge a' = 0$.

\par
\medskip
\noindent
This gives us the next axiom:

\begin{axiom} [orthocomplementation]
A state property system $(\Sigma, {\cal L}, \xi,
\kappa)$ describing an entity $S$ is called `property
orthocomplemented' and satisfies axiom 4 iff there exists an
orthocomplementation on the complete lattice of properties.
\end{axiom}
Apart from the orthocomplementation the property lattice of an entity
described by standard quantum mechanics satisfies an additional
property called `weak modularity'. It is a purely technical axiom
expressed as follows:

\begin{axiom} [weak modularity]
The property lattice ${\cal L}$ of a state property space $(\Sigma,
{\cal L}, \xi, \kappa)$ satisfying axiom 1 and 4 (hence
a property orthocomplemented state property system) is `weakly
modular' iff for $a, b \in {\cal L}$ such that $a \prec b$ we have $(a
\vee b')
\wedge b = a$.
\end{axiom}

\subsection{Full axiomatisation of standard quantum mechanics}

\noindent
We need more requirements in order to be able to prove that
the obtained structure is isomorphic to standard
quantum mechanics. We leave the proof for these requirements to be
satisfied in standard quantum mechanics to the dedicated reader. The
first requirement is called `plane transitivity'. It has been
identified only recently ([32] [33]).

\begin{axiom} [plane transitivity]
The property lattice ${\cal L}$ of a state property space $(\Sigma,
{\cal L}, \xi, \kappa)$ is `plane transitive' iff for $p, q \in
\Sigma$ there are $r \not= s \in \Sigma$ and an automorphism
of ${\cal L}$ that maps $p$ onto $q$ and leaves the
`plane' interval $[0,r \vee s]$ invariant.
\end{axiom}

Let us introduce the next axiom:
\begin{axiom} [irreducibility]
The property lattice ${\cal L}$ of a state property space $(\Sigma,
{\cal L}, \xi, \kappa)$ satisfying axiom 1 and 2 (hence
a property orthocomplemented state property system) is `irreducible'
i.e. whenever
$b
\in {\cal L}$ is such that $b = (b \wedge a) \vee (b \wedge a')\
\forall\ a \in {\cal L}$ then $b = 0$ or $b = I$.
\end{axiom}

The standard representation theorem has been proven for the
irreducible components of the property lattice. The foregoing axiom is
in this sense not on the same level as the other ones. Indeed even if
we do not require the property lattice to be irreducible, the
representation theorem can be proven for each irreducible component.
It can indeed be shown that a general property lattice is the direct
product of its irreducible components. We refer to [38] and [41]
and more specifically to [42] and [43] for a detailed analysis of this
decomposition.

When we mentioned the representation theorem derived from the
fundamental theorem of projective geometry in section \ref{linear} we
already pointed out that the property lattice has to contain enough
states to be able to derive this theorem. For the full
representation theorem of standard quantum mechanics we need
infinitely many atoms.

\begin{axiom} [infinite length]
The property lattice ${\cal L}$ of a state property space $(\Sigma,
{\cal L}, \xi, \kappa)$ is `infinite' if it contains an infinite set
of mutually orthogonal elements\footnote{Two elements are orthogonal
iff they imply respectively a property and its orthocomplement.}.
\end{axiom}

\begin{theorem} [Representation theorem]
Suppose that we have an entity $S$ described by means of a state
property space $(\Sigma, {\cal L}, \xi, \kappa)$ for which axioms 1 to
8 are satisfied. Then ${\cal L}$ is isomorphic to the complete lattice
of the projection operators of an infinite dimensional real, complex
or quaternionic Hilbert space
${\cal H}$. The atoms of ${\cal L}$ and hence also the atomic states
of $S$ correspond to the rays of ${\cal H}$. The orthocomplementation
is induced by the orthogonality structure of ${\cal H}$.
\end{theorem}
We will not prove this theorem, but refer to [32, 33, 34, 35,
36, 37, 38, 39, 40, 41, 42, 43, 44, 45, 46, 47, 48, 49] where pieces
preparing the proof can be found. We refer to [32, 33] for
a recent and more complete overview and the inclusion of the new
axiom of plane transitivity.

\section{Paradoxes and failing axioms} \label{failingaxioms}

\noindent
The aim of quantum axiomatics was to construct an operational
foundation for standard quantum mechanics starting from basic
concepts, states and properties, that are easy to identify
physically. Once a full axiomatics has been constructed this gives of
course a powerful tool to investigate the well known paradoxes that
quantum mechanics entails. Let us investigate some aspects of this
possibility.

\subsection{The description of entity consisting of two entities}

\noindent
We will consider the description of two spins by using the
quantum machine model for these spins. So we consider now
two quantum machines. Let us call them $S_1$ and $S_2$, and the entity
that just consists of these two quantum machines. In a general way, the
entity $S_1$ is described by a state property space $(\Sigma_1, {\cal
L}_1, \xi_1, \kappa_1)$ and the entity $S_2$ is described by a state
property space
$(\Sigma_2, {\cal L}_2, \xi_2, \kappa_2)$. Let us denote the sphere
of the first quantum machine $S_1$ by $ball_1$, the points of this
sphere by $v_1 \in ball_1$, and the states, the experiments
and the properties connected to this quantum machine by $p_{v_1}$,
$e_{u_1}$ and $a_{u_1}$. In an analogous way we denote the sphere
of the second quantum machine $S_2$ by $ball_2$, and its states,
experiments and properties by $p_{v_2}$, $e_{u_2}$ and $a_{u_2}$.

As we have shown in (\ref{propspin}) and (\ref{statspin}) the sets of
properties, the sets of states and the sets of experiments are given
by:
\begin{equation}
\begin{array}{l}
{\cal L}_1 = \{a_{u_1}\ \vert\ u_1 \in ball_1, |u_1| = 1\} \\
\Sigma_1 = \{p_{v_1}\ \vert\ v_1 \in ball_1\} \\
{\cal E}_1 = \{e_{u_1}\ \vert\ u_1 \in ball_1, |u_1| = 1\} \\
{\cal L}_2 = \{a_{u_2}\ \vert\ u_2 \in ball_2, |u_2| = 1\} \\
\Sigma_2 = \{p_{v_2}\ \vert\ v_2 \in ball_2\} \\
{\cal E}_2 = \{e_{u_2}\ \vert\ u_2 \in ball_2, |u_2| = 1\}
\end{array}
\end{equation} 
Let us
call $S$ the compound physical entity, consisting of the two quantum
machines, and $(\Sigma, {\cal L}, \xi, \kappa)$ the state property
space describing this entity $S$. To see in which way the three state
property spaces are connected we have to analyse the physical
situation.

\par
\medskip
\noindent
{\it The states}

\par
\medskip
\noindent
Clearly a state $p$ of the entity $S$ completely determines a
state $p_1$ of $S_1$ and a state $p_2$ of $S_2$ - using the physical
principle that when the entity $S$ `is' in state $p$ then the
entities $S_1$ and $S_2$ `are' in two corresponding states $p_1$ and
$p_2$\footnote{If we take the ontological meaning of the concept of
state seriously, we can hardly ignore this physical principle.
Although, as we will see, standard quantum mechanics gives rise to
problems here.}. This defines two functions:
\begin{equation}
\begin{array}{ll} \label{coup01}
m_1 : \Sigma \rightarrow \Sigma_1 & p \mapsto m_1(p) \\
m_2 : \Sigma \rightarrow \Sigma_2 & p \mapsto m_2(p)
\end{array}
\end{equation}

\par
\medskip
\noindent
{\it The properties}

\par
\medskip
\noindent
Each experiment $e_1$ on $S_1$ is also an experiment on $S$ and each
experiment $e_2$ on $S_2$ is also an experiment on $S$ - following
the physical principle that if we perform an experiment on one of the
sub-entities we perform it also on the compound
entity\footnote{Again, if we take the meaning of what an experiment
is seriously, it is hard to ignore this principle. We have even no
reason here to doubt it, because standard quantum mechanics
agrees with it.}. Since the properties are operationally defined by means 
of the
experiments, from the same physical principle, we have that each
property of a sub-entity is also a property of the compound entity.
This defines again two functions:
\begin{equation}
\begin{array}{ll} \label{coup02}
n_1 : {\cal L}_1 \rightarrow {\cal L} & a_1 \mapsto n_1(a_1) \\
n_2 : {\cal L}_2 \rightarrow {\cal L} & a_2 \mapsto n_2(a_2)
\end{array}
\end{equation}

\par
\medskip
\noindent
{\it A covariance principle}

\par
\medskip
\noindent
If property $a_1$ is actual for entity $S_1$ in state $m_1(p)$, then
property $n_1(a_1)$ is actual for entity $S$ in state $p$. An
analogous covariance principle is satisfied between entity $S_2$ and
entity $S$. This means that we have the following equations:

\begin{equation}
\begin{array}{l} \label{coup03}
a_1 \in \xi_1(m_1(p)) \Leftrightarrow n_1(a_1) \in \xi(p) \\
a_2 \in \xi_2(m_2(p)) \Leftrightarrow n_2(a_1) \in \xi(p)
\end{array}
\end{equation}
It has been shown in [50], [30], [31] and [33], that for the case of
physical entities satisfying axiom 1 (hence the three state property
spaces are state property systems) this covariance principle gives rise
to a unique minimal structure for the state property system of the
compound entity. It is the (co)product in the category of state
property systems.

\begin{theorem} \label{coprod}
Suppose that we have two entities $S_1$ and $S_2$ with state
property systems
$(\Sigma_1, {\cal L}_1,
\xi_1,
\kappa_1)$ and $(\Sigma_2, {\cal L}_2, \xi_2, \kappa_2)$ that form a
compound entity $S$ with a state property system $(\Sigma, {\cal L},
\xi, \kappa)$ according to (\ref{coup01}), (\ref{coup02}) and
(\ref{coup03}). The minimal solution is as follows:
\begin{equation}
\Sigma = \Sigma_1 \times \Sigma_2
\end{equation}
where $\Sigma_1 \times \Sigma_2$ is the cartesian product of
$\Sigma_1$ and $\Sigma_2$. For $(p_1,p_2) \in \Sigma$ we have:
\begin{equation}
\begin{array}{ll}
m_1(p_1,p_2) = p_1 & m_2(p_1,p_2) = p_2
\end{array}
\end{equation} 
For $p_1, q_1 \in \Sigma_1$ and $p_2, q_2 \in \Sigma_2$ we have:
\begin{equation}
(p_1, p_2) \prec (q_1, q_2) \Leftrightarrow p_1 \prec q_1\ {\rm and}\
p_2 \prec q_2
\end{equation}
\begin{eqnarray}
{\cal L} &=& {\cal L}_1 \coprod {\cal L}_2 \\
&=& \{(a_1,a_2)\ \vert\ a_1 \in {\cal L}_1, a_2 \in {\cal L}_2, \\
\nonumber
 & & a_1
\not= 0_1, a_2 \not=0_2\} \cup \{0\}
\end{eqnarray}
where ${\cal L}_1 \coprod {\cal L}_2$ is called the co-product of
${\cal L}_1$ and ${\cal L}_2$. For
$a_1
\in {\cal L}_1$ and
$a_2 \in {\cal L}_2$ we have:
\begin{equation}
\begin{array}{l}
n_1(a_1) = (a_1, I_2)\ {\rm if}\ a_1 \not= 0_1 \\
n_1(0_1) = 0 \\
n_2(a_2) = (I_1, a_2)\ {\rm if}\ a_2 \not= 0_2 \\
n_2(0_2) = 0
\end{array}
\end{equation}
For $a_1, b_1, a^i_1 \in {\cal L}_1$ and $a_2, b_2, b^i_2 \in {\cal
L}_2$ we have:
\begin{equation}
(a_1,a_2) \prec (b_1, b_2) \Leftrightarrow a_1 \prec b_1\ {\rm and}\
a_2 \prec b_2
\end{equation}
\begin{equation}
0 \prec (a_1, a_2)
\end{equation}
\begin{eqnarray}
\wedge_i(a^i_1,a^i_2) &=& (\wedge_ia^i_1, \wedge_ia^i_2) \\
\nonumber
& & {\rm if}\ \wedge_ia^i_1 \not=0_1\ {\rm and}\ \wedge_ia^i_2
\not= 0_2 \\ &=& 0\ \ \ {\rm if}\  \wedge_ia^i_1 = 0_1\ {\rm or}\
\wedge_ia^i_2 = 0_2
\end{eqnarray}
Further we have:
\begin{equation}
\xi(p_1,p_2) = \{(a_1,a_2)\ \vert\ a_1 \in \xi_1(p_1), a_2 \in
\xi_2(p_2)\}
\end{equation}
\begin{equation}
\kappa(a_1,a_2) = \{(p_1,p_2)\ \vert\ p_1 \in \kappa(a_1), p_2 \in
\kappa(a_2)\}
\end{equation}
\end{theorem}

\subsection{The covering law and compound entities}

\noindent
This structure of (co)product is the simplest one that can be
constructed for the description of the compound physical entity $S$.
One would expect that it is `the' structure to be used to describe the
compound entity $S$. This is however not the case for quantum
entities in standard quantum mechanics. The reason is that the
co-product `never' satisfies two of the axioms of standard quantum
mechanics, namely axiom 3 (covering law) and axiom 4
(orthocomplementation). Let us prove this for the case of the covering
law.

\begin{theorem}
Suppose that axiom 1  and 2 are satisfied and consider two entities
$S_1$ and $S_2$ described by state property systems $(\Sigma_1, {\cal
L}_1, \xi_1, \kappa_1)$ and $(\Sigma_2, {\cal L}_2, \xi_2, \kappa_2)$
and the minimal compound entity $S$ consisting of $S_1$ and $S_2$ and
described by the state property system $(\Sigma_1 \times \Sigma_2,
{\cal L}_1 \coprod {\cal L}_2, \xi,\kappa)$ as defined in theorem
\ref{coprod}. Suppose that axiom 3 is satisfied for the entity $S$,
then one of the two entities $S_1$ or $S_2$ has a trivial property
lattice consisting only of the minimal and maximal element.
\end{theorem}
Proof: Suppose that ${\cal L}_1$ has at least one element $a_1$
different from $I_1$ and $0_1$. Since ${\cal L}_1$ is atomistic there
exists an atom $c_1 \prec a_1$, and at least one atom $d_1 \not\prec
a_1$. Hence $c_1 \not= d_1$. Consider now two arbitrary atoms
$c_2, d_2 \in {\cal L}_2$. We have:
\begin{equation}
(c_1,c_2) \prec (c_1 \vee d_1, c_2) \prec (c_1 \vee d_1, c_2 \vee
d_2)
\end{equation}
\begin{equation}
(c_1 \vee d_1, c_2 \vee
d_2) = (c_1,c_2) \vee (d_1, d_2)
\end{equation}
Since $(d_1, d_2)$ is an atom of ${\cal L}_1 \coprod {\cal L}_2$, the
property $(c_1,c_2) \vee (d_1, d_2)$ `covers' $(c_1,c_2)$, because the
covering law is satisfied for $S$. We therefore have:
\begin{equation}
\begin{array}{l}
(c_1,c_2) = (c_1 \vee d_1, c_2)\ {\rm or} \\
 (c_1 \vee d_1, c_2) = (c_1 \vee d_1, c_2 \vee d_2)
\end{array}
\end{equation}
This implies that
\begin{equation}
c_1 = c_1 \vee d_1\ \ {\rm or}\ \ c_2 = c_2 \vee d_2
\end{equation}
Since $c_1 \not= d_1$ we cannot have $c_1 = c_1 \vee d_1$. Hence we
have $c_2 = d_2$. Since $c_2$ and $d_2$ were arbitrary atoms of ${\cal
L}_2$, this proves that ${\cal L}_2$ contains only one atom. From this
follows that ${\cal L}_2 = \{0_2,I_2\}$.

\par
\medskip
\noindent
This theorem proves that for two non-trivial entities $S_1$ and $S_2$
the property lattice that normally should represent the compound entity
$S$ never satisfies the covering law. This same theorem also proves
that, since we know that for an entity described by standard quantum
mechanics the covering law is satisfied for its property lattice, in
quantum mechanics the compound entity $S$ is `not' described by the
minimal product structure. The covering law, as we remarked earlier
already, is the axiom that introduces the linear structure for the
state space. This means that for a property lattice that does not
satisfy the covering law it will be impossible to find a vector space
representation such that the superposition principle of standard
quantum mechanics is available.
It can be shown that this fact is at the
origin of the Einstein Podolsky Rosen paradox as it is
encountered in quantum mechanics (see [19] and [20]). It means indeed
that the compound entity $S$, as it is described in standard quantum
mechanics, will have additional elements in its state property
structure, that are not contained in the minimal product structure
that we have proposed here. As we will see, these additional elements
are the so called `non-product states'.

\subsection{The quantum description of the compound entity}

\noindent
For standard quantum mechanics the compound entity $S$ consisting of
two entities $S_1$ and $S_2$ is described by means of the tensor
product ${\cal H}_1 \otimes {\cal H}_2$ of the Hilbert spaces ${\cal
H}_1$ and ${\cal H}_2$ that describe the two sub-entities $S_1$ and
$S_2$. We have studied this situation in detail in earlier
work [51], and will here only expose the scheme. 

Let us consider an entity $S$ described with a state property
space $(\Sigma,
{\cal L}, \xi, \kappa)$ corresponding to the
Hilbert space ${\cal H}$ consisting of two entities $S_1$ and $S_2$
described by state property spaces $(\Sigma_1, {\cal L}_1, \xi_1,
\kappa_1)$ and $(\Sigma_2, {\cal L}_2, \xi_2, \kappa_2)$
corresponding to Hilbert space ${\cal H}_1$ and ${\cal H}_2$. Let us
first identify the functions $m$ and $n$ that describe the situation
where $S$ is the joint entity of $S_1$ and $S_2$. For the function $n$
this identification is straightforward. We have:
\begin{equation}
\begin{array}{ll}
n_1: {\cal L}_1 \rightarrow {\cal L} &
a_{P_1} \mapsto a_{P_1 \otimes I_2} = n_1(a_{P_1}) \\ 
n_2: {\cal L}_2 \rightarrow {\cal L} &
a_{P_2} \mapsto a_{I_1 \otimes P_2} = n_2(a_{P_2}) 
\end{array}
\end{equation}
This shows that for standard quantum mechanics, as in the case where
we would describe the compound entity by means of the co-product, for
each property $a_1$ of $S_1$ ($a_2$ of $S_2$) there is a unique
property $n_1(a_1)$ ($n_2(a_2)$) of $S$. The requirement that with
each state $p$ of the compound  entity
$S$ correspond unique states $p_1$ and $p_2$ of the sub-entities gives
rise to a special situation in the case that $p$ corresponds to a
non-product vector of the Hilbert space ${\cal H}_1
\otimes {\cal H}_2$, i.e. a vector $c = \sum_ic^i_1 \otimes c^i_2$
that cannot be reduced to a product of a vector in ${\cal H}_1$ and
a vector in ${\cal H}_2$. It can be shown that, taking into account
the covariance requirement - this time also for the probalitities -
there do correspond two unique states $p_{W_1}$ and $p_{W_2}$ to
such a state $p_{\bar c}$, but when $c$ is a non-product vector,
$W_1$ and $W_2$ are density operators that do not correspond to
vectors in ${\cal H}_1$ and ${\cal H}_2$. This means (see
theorem \ref{densatom}) that $p_{W_1}$ and $p_{W_2}$ are non atomic
states although $p_{\bar c}$ is an atomic state. We have not stated
this too explicit till now, but in standard quantum mechanics there
is a real physical difference between the atomic states - represented
by density operators corresponding to a vector - and the non atomic
states - represented by density operators not corresponding to a
vector. The atomic states are `pure states' and the non atomic states
are `mixed states'. This is in fact also the case in our operational
definition of the join states in (see definition
\ref{productteststate}). The join state of a set of states, as defined
there is a mixture of these states, which means that the entity is
in the join state of this set iff it is in one of the states of this
set, but we lack the knowledge about which one. So we repeat: a
mixed state of a set of pure states describes our lack of knowledge
about the pure state where the entity is in. If this is the meaning of
a non atomic state, hence a mixed state over some set of atomic
states, as it is the case in standard quantum mechanics, we can
conclude that the entity is always in an atomic state. The non atomic
states only describe our lack of knowledge about the atomic state the
entity exactly is in.

We can now see where the fundamental problem arises with the tensor
product coupling procedure of quantum mechanics. If entities are
always in atomic states, and since for an atomic state of the
compound entity that corresponds to a non-product vector of the
tensor product Hilbert space, the component states are strictly non
atomic, it would indicate that the sub-entities are not in a
state. This is of course very strange. Indeed, it seems even
contradictory with the concept of state itself. An entity must always
be in a state (and hence a quantum entity always in an atomic state),
whether it is a sub-entity of another entity or not.

\par
As we have
mentioned already, these non-product states also give rise
to EPR type correlations between the two sub-entities $S_1$ and $S_2$.
We remark that the presence of these correlations also indicates that
the quantum description of the compound entity is not a description of
`separated' entities. So something really profoundly mysterious occurs
here. We also mention that it is excluded that the non-product states
of the quantum compound entity would be mathematical artifacts of the
theory, since entities are without many problems prepared in these
non-product states in the laboratory these days\footnote{The question
about the reality of the non-product states was settled during the
second half of the seventies and the first half of the eighties by
means of the well known Einstein Podolsky Rosen correlation
experiments. Meanwhile it has become common laboratory practice to
prepare `entangled' entities  - that is what they are referred to now
in the literature - in non-product states.}. So the non-product states
exists and are real states of the compound entity consisting of two
quantum entities. Should we then decide that the sub-entities have
disappeared as entities, and only some properties are left? We want to
reflect more about this question and investigate what the
possibilities are. Most of all we want to put forward an alternative
possibility, that is however speculative, but should be worth further
investigation.

\subsection{About mixtures, pure states non atomic pure states}

\noindent
We have to remark that the problem that we explained in the foregoing
section was known from the early days  of quantum mechanics but
concealed more or less by the confusion that often exists between pure
states and mixtures. Let us explain this first. The reality of a
quantum entity in standard quantum mechanics is represented by a pure
state, namely a ray of the corresponding Hilbert space.
Mixed states are represented in standard quantum
mechanics by density operators (positive self adjoint operators with
trace equal to 1). But although a mixed state is also called a state,
it does not represent the reality of the entity under consideration,
but a lack of knowledge about this reality. This means that if the
entity is in a mixed state, it is actually in a pure state, and the
mixed state just describes the lack of knowledge that we have about
the pure state it is in. We have remarked that the deep conceptual
problem that we indicate here was noticed already in the early days of
quantum mechanics, but disguised by the existence of the two types of
states, pure states and mixed states. Indeed in most books on quantum
mechanics it is mentioned that for the description of sub-entities by
means of the tensor product procedure it is so that the compound entity
can be in a pure state (and a non-product state is meant here) such
that the sub-entities will be in mixed states and not in pure states
(see for example [52] 11-8 and [53] p 306). The fact that the
sub-entities, although they are not in a pure state, are at least in a
mixed state, seems at first sight to be some kind of a solution to the
conceptual problem that we indicated in the foregoing section.
Although a little further reflection shows that it is not: indeed, if
a sub-entity is in a mixed state, it should anyhow be in a pure state,
and this mixed state should just describe our lack of knowledge about
this pure state. So the problem is not solved at all. Probably because
quantum mechanics is anyhow entailed with a lot of paradoxes and
mysteries, the problem was just added to the list by the majority of
physicists.

Way back, in a paper published in 1984, we have already shown that in
a more general approach we can  define pure states for the
sub-entities, but they will not be `atoms' of the lattice of properties
[50]. As we have shown already the ray states of quantum mechanics
give rise to atoms of the property lattice, such that `pure states' in
quantum mechanics correspond to `atomic' states of the state property
space. This means that the non atomic pure states that we have
identified in [50] can anyhow not be represented within the standard
quantum mechanical formalism. We must admit that the finding of the
existence of non atomic pure states in the 1984 paper, even from the
point of view of generalized quantum formalisms, seemed also to us
very far reaching and difficult to interpret physically. Indeed
intuitively it seems to be so that only atomic states should represent
pure states. We know now that this is a wrong intuition. But to
explain why we have to present first the other pieces of the puzzle.

A second piece of the puzzle appeared when in 1990 we built a model of
a mechanistic classical laboratory situation violating Bell inequalities
with $\sqrt{2}$, exactly `in the same way' as its violations by the EPR 
experiments [54]. With this model we tried to show that Bell
inequalities can be violated in the macroscopic world with the same
numerical value as the quantum violation. What is interesting for
the problem of the description of sub-entities is that new `pure'
states were introduced in this model. We will see in a moment that the
possibility of existence of these new states lead to a
possible solution of the problem of the description of sub-entities
within a Hilbert space setting, but different from standard quantum
mechanics.

More pieces of the puzzle appeared steadily during the elaboration of 
the general formalism presented in [6]. We started to
work on this formalism during the first half of the
eighties, reformulating and elaborating some of the concepts during
these years. Then it became clear that the new states introduced in
[54], although `pure' states in the model, appear
as non atomic states in the general formalism. This made us understand
that the first intuition that classified non atomic states as bad
candidates for pure states was a wrong intuition. Let us present now the
total scheme of our possible solution.

In the example that we proposed in [54] we used two  spin
models as the one presented here (the quantum machine) and introduced
new states on both models with the aim of presenting a situation that
violates the Bell inequalities exactly as in the case of the singlet
spin state of two coupled spin${1 \over 2}$ particles do. We indeed
introduced a state for both spin models that corresponds to the point
in the center of each sphere, and connecting these two states by a
rigid rod we could generate a violation of Bell's inequalities. We
have shown in the last part of section \ref{complquant} that the centre
of $ball$ is a non atomic state of the quantum machine. This
means that we have `identified' a possible `non-mixture' state
(meaning with `non-mixture' that it really represents the reality of
the entity and not a lack of knowledge about this reality) that is not
an atom of the pre-ordered set of states. Is this a candidate for the
`non-mixed' states that we identified in [50] and that were
non-atoms? It is indeed, as we prove explicitly in [6]. The states
$p_w$, where $w \in ball$ and $|w| < 1$, as defined in
\ref{complquant}, and that are certainly pure states for the quantum
machine entity, are represented by the density operator
$W(w)$ if we use formula~\ref{probcal} for the calculation of the
transition probabilities.

\subsection{Completing quantum mechanics?}

\noindent
The idea that we want to put forward is the following: perhaps
density operators just do represent pure states, also for a quantum
entity. Such that the set of pure states would be represented by the
set of density operators and not by the set of rays of the Hilbert
space. If this would be the case, the conceptual problem of using the
tensor product would partly be solved. We admit immediately that it
is a very speculative idea that we put forward here. The problem is
also that it will be difficult to test it experimentally on one
physical entity. Indeed, the so called new pure states, corresponding
to density operators of the Hilbert space, cannot experimentally be
distinguished from mixtures of old pure states,
corresponding to vectors, and represented also by these density
operators. It is an easy mathematical result since in all
probability calculations only the density operator appears. We can
also see it explicitly on the quantum machine. Let us go back to the
calculation that we made in the second part of section
\ref{complquant}. The `pure' state $p_w$ corresponding to an interior
point
$w = \lambda_1 \cdot v + \lambda_2 \cdot (-v)$ of $ball$ is represented
by the density operator 
\small
\begin{equation} \label{densop}
W(w) = \left( \begin{array}{ccc} 
\lambda_1\cos^2{\theta \over 2} + \lambda_2\sin^2{\theta \over 2} & 
(\lambda_1-\lambda_2)\sin{\theta
\over 2}\cos{\theta \over 2}e^{-i\phi}  \\
(\lambda_1-\lambda_2)\sin{\theta
\over 2}\cos{\theta \over 2}e^{i\phi} & \lambda_1\sin^2{\theta \over 2} +
\lambda_2\cos^2{\theta
\over 2} 
\end{array} \right)
\end{equation}
\normalsize
But this density operator represents also the `mixed' state describing
the following situation of lack of knowledge: the point is in one of
the pure states $p_v$ or $p_{-v}$ with weight $\lambda_1$ and
$\lambda_2$ respectively. Although this mixed state and the pure
state $p_w$ are ontologically very different states, they cannot
experimentally be distinguished by means of the elastic experiments.

\section{From Euclid to Riemann: the quantum \\ mechanical equivalent}

\noindent
We have called section \ref{failingaxioms} `Paradoxes and failing
axioms'. Indeed a possible conclusion for the result
that we have exposed in this section is that the covering law (and
some of the other axioms) are just no good and that we should look for
a more general formalism than standard quantum mechanics. We are more
and more convinced that this must be the case, also because it is again
the covering law that makes it impossible to describe a continuous
change from quantum to classical passing through the intermediate
situations that we have mentioned in section \ref{intermediate} (see
[5], [25] and [27] and the paper of Thomas Durt in this book).

In this sense we want to come back now to the suggestive idea that was
proposed in the introduction. The idea of relativity theory, to
take the points of space-time as representing the events of reality,
goes back to a long tradition. It was Euclid who for the first time
synthesised the descriptions that the Greek knew about the properties
of space by using as basic concepts points, lines and planes. Let us
remember that Euclid constructed an axiomatic system consisting of
five axioms - now called
Euclid's axioms - for Euclidean space and its geometric
structure. All classical physical theories have later, without
hesitation, taken the Euclidean space as theatre for reality. From a
purely axiomatic point of view there has been a long and historic
discussion about the independence of the fifth Euclidean axiom. Some
scientists have pretended to be able to derive it from the four other
axioms. The problem was resolved in favour of the independence by
Gauss, Bolyai and Lobachevski by constructing explicit models of
non-euclidean geometries. Felix Klein proposed a classification along
three fundamental types: an elliptic geometry, the one originally
proposed by Gauss, the geometry of the surface of a sphere for
example, a hyperbolic geometry, proposed by Lobachevski, the
geometry of the surface of a saddle for example, and a parabolic
geometry, that lies in between both. It was Riemann who proposed a
complete theory of non-euclidean geometries, the geometry of curved
space, that was later used by Einstein to formulate general
relativity theory. It is interesting to remark that the programme of
general relativity - to introduce the force fields of physics as
properties of the metric of space - was already put forward
explicitly by Riemann. He could however not have found the solution
of general relativity because he was looking for a solution in three
dimensional space, and general relativity has to be constructed in
four dimensional space-time. In the formulation of Einstein, which is
the one of Riemann applied to the four dimensional space-time
continuum, the events are represented by the points and the metric
tensor describes the nature of the geometry. Let us now see whether
we can find an analogy with our analysis of quantum mechanics by
means of our axiomatic approach. Here the basic concepts are
states, properties and probability. In geometry a set of points
forms a space. In quantum mechanics the set of states and properties
form a state property space. Just like the Euclidean space is not just
any space, the state property space of quantum mechanics is not just
any state property space. We have outlined $8$ axioms that make an
arbitrary state property space into a quantum mechanical state
property space. We have remarked already that the purpose of quantum
axiomatics was not to change standard quantum mechanics, exactly as
the purpose of Euclid was certainly not to formulate an alternative
geometry: he wanted to construct an operational theory about the
structure of space. 

\par
The operational axiomatization of quantum theory
has taken a long time, the axiom of `plane transitivity' relies
on a result of Maria Pia Soler of 1995 [48]. Since however the general
problems related to axiomatization are nowadays better known, there
has been little discussion about the independence of the axioms. But
the physical meaning of certain axioms (e.g. the failing covering law)
remained obscure. These were merely axioms introduced to recover
the complete Hilbert space structure which includes the linearity
of the state space. Within this development of the axiomatic
structure of quantum theory it has been shown now that the compound
entity of two separated quantum entities cannot be described. The
axiom that makes such a description impossible is the covering law,
equivalent to the linearity of the state space. Do we have to work
out a general quantum theory, not necessarily satisfying the covering
law, as Riemann has given us a general theory of space? We believe
so, but we know that a lot of work has to be done. In [6] a
humble general scheme is put forward that may be a start for the
elaboration of such a general quantum theory. 

\par
\bigskip
\noindent
I want to thank Jan Broekaert and Bart Van Steirteghem to read and
discuss with me parts of the text of this article. Their remarks and
suggestions have been of great value.

\section{References}

\noindent
[1] {\bf Von Neumann, J.},{\it Mathematische Grundlagen
der Quanten-Mechanik}, Springer-Verlag, Berlin, 1932.

\smallskip
\noindent
[2] {\bf Birkhoff, G. and Von Neumann, J.}, ``The logic of
quantum mechanics", {\it Annals of Mathematics}, {\bf 37}, 823, 1936.

\smallskip
\noindent
[3] {\bf Foulis, D.}, ``A half-century of quantum logic - what have we
learned?" in {\it Quantum Structures and the Nature of Reality}, the {\it 
Indigo
Book of Einstein meets Magritte}, eds.,{\bf Aerts, D. and Pykacz, J.},
Kluwer Academic, Dordrecht, 1999.

\smallskip
\noindent
[4] {\bf Jammer, M.}, {\it The Philosophy of Quantum Mechanics}, Wiley 
and
Sons, New York, Sydney, Toronto, 1974.

\smallskip
\noindent
[5] {\bf Aerts, D. and Durt, T.}, ``Quantum, classical and intermediate, 
an
illustrative example", {\it Found. Phys.\/} {\bf 24}, 1353, 1994.

\smallskip
\noindent
[6] {\bf Aerts, D.}, ``Foundations of physics: a general realistic and
operational approach", {\it Int. J. Theor. Phys.}, {\bf 38}, 289, 1999, and quant-ph/0105109.

\smallskip
\noindent
[7] {\bf Aerts, D.}, ``A possible explanation for the probabilities of
quantum mechanics and example of a macroscopic system that violates Bell
inequalities", in {\it Recent developments in quantum logic}, eds., 
{\bf
Mittelstaedt, P. and Stachow, E.W.}, {\it Grundlagen der Exakten
Naturwissenschaften, band 6, Wissenschaftverlag}, Bibliografisches 
Institut,
Mannheim, 1985.

\smallskip
\noindent
[8] {\bf Aerts, D.}, ``A possible explanation for the probabilities of
quantum mechanics", {\it J. Math. Phys.\/} {\bf 27}, 202, 1986.

\smallskip
\noindent
[9] {\bf Aerts, D.}, ``Quantum structures : an attempt to explain their
appearance in nature", {\it Int. J. Theor. Phys.\/} {\bf 34}, 1165, 1995.

\smallskip
\noindent
[10] {\bf Gerlach, F. and Stern, O.}, ``Der experimentelle Nachweis der
Richtungsquantelung im Magnetfeld", Zeitschrift f{\"u}r Physik {\bf
9}, 349, 1922.

\smallskip
\noindent
[11] {\bf Bell, J. S.}, {\it Rev. Mod. Phys.}, {\bf 38}, 447, 1966.

\smallskip
\noindent
[12] {\bf Jauch, J. M. and Piron, C.}, {\it Helv. Phys. Acta}, {\bf 36},
827, 1963.

\smallskip
\noindent
[13] {\bf Gleason, A. M.}, {\it J. Math. Mech.}, {\bf 6}, 885, 1957.

\smallskip
\noindent
[14] {\bf Kochen, S. and Specker, E. P.}, {\it J. Math. Mech.}, {\bf 17},
59, 1967.

\smallskip
\noindent
[15] {\bf Gudder, S. P.}, {\it Rev. Mod. Phys.}, {\bf 40}, 229, 1968.

\smallskip
\noindent
[16] {\bf Accardi, L.}, {\it Rend. Sem. Mat. Univ. Politech. Torino}, 
241, 1982.

\smallskip
\noindent
[17] {\bf Accardi, L. and Fedullo, A.}, {\it Lett. Nuovo Cimento}, {\bf 
34},
161, 1982. 

\smallskip
\noindent
[18] {\bf Aerts, D.},  ``Example of a macroscopical situation that 
violates
Bell inequalities", {\it Lett. Nuovo Cimento}, {\bf 34}, 107, 1982,

\smallskip
\noindent
[19] {\bf Aerts, D.}, ``The physical origin of the EPR paradox", in {\it
Open questions in quantum physics}, eds., {\bf Tarozzi, G. and van der 
Merwe,
A.}, Reidel, Dordrecht, 1985.

\smallskip
\noindent
[20] {\bf Aerts, D.}, ``The physical origin of the Einstein-Podolsky-
Rosen
paradox and how to violate the Bell inequalities by macroscopic systems", 
in {\it
Proceedings of the Symposium on the Foundations of Modern Physics}, 
eds., {\bf Lahti, P. and Mittelstaedt, P.} World Scientific, Singapore, 
1985.

\smallskip
\noindent
[21] {\bf Aerts, D.}, ``Quantum structures, separated physical entities 
and
probability", {\it Found. Phys.\/} {\bf 24}, 1227, 1994.

\smallskip
\noindent 
[22] {\bf Coecke, B.,} ``Hidden measurement representation for quantum 
entities described by finite dimensional complex Hilbert spaces", {\it 
Found. Phys.}, {\bf 25}, 203, 1995.

\smallskip
\noindent
[23] {\bf Coecke, B.,} ``Generalization of the proof on the existence of 
hidden measurements to experiments with an infinite set of outcomes", 
{\it Found. Phys. Lett.}, {\bf 8}, 437, 1995.

\smallskip
\noindent
[24] {\bf Coecke, B.,} ``New examples of hidden measurement systems and 
outline of a general scheme", {\it Tatra Mountains Mathematical 
Publications}, {\bf 10}, 203, 1996.

\smallskip
\noindent
[25] {\bf Aerts, D. and Durt, T.,} ``Quantum, classical and intermediate: 
a measurement model", in
the proceedings of the {\it International Symposium on the Foundations of
Modern Physics 1994, Helsinki, Finland}, eds., {\bf Montonen, C., et al.}, Editions
Frontieres, Gives Sur Yvettes, France, 1994. 

\smallskip
\noindent
[26] {\bf Aerts, D., Durt, T. and Van Bogaert, B.}, ``A physical example
of quantum fuzzy sets, and the classical limit", {\it Tatra Mt. Math. Publ.}, 5 - 15, 1993.

\smallskip
\noindent
[27] {\bf Aerts, D., Durt, T. and Van Bogaert, B.}, ``Quantum 
probability,
the classical limit and non-locality", in {\it the proceedings of the
International Symposium on the Foundations of Modern Physics 1992}, 
Helsinki,
Finland, ed. {\bf Hyvonen, T.}, World Scientific, Singapore, 35, 1993.  

\smallskip
\noindent
[28] {\bf Randall, C. and Foulis, D.}, `` Properties and
operational propositions in quantum mechanics", {\it  Found. Phys.},
{\bf 13}, 835, 1983.

\smallskip
\noindent
[29] {\bf Foulis, D., Piron, C. and Randall, C.}, ``Realism,
operationalism, and quantum mechanics", {\it Found. Phys.}, {\bf 13}, 
813, 1983.

\smallskip
\noindent
[30] {\bf Aerts, D., Colebunders, E., Van der Voorde, A. and Van
Steirteghem, B.}, ``State property systems and closure spaces: a study of
categorical equivalence", {\it Int. J. Theor. Phys.}, {\bf 38}, 259, 1999, and quant-ph/0105108.

\smallskip
\noindent
[31] {\bf Aerts, D., Colebunders, E., Van de Voorde, A., and Van Steirteghem, B.}, ``The construct of closure spaces as the amnestic
modification of the physical theory of state property systems",  accepted for publication in {\it Applied Categorical Structures}.

\smallskip
\noindent
[32] {\bf Aerts, D. and Van Steirteghem, B.}, ``Quantum axiomatics and a
theorem of M.P.~Sol\`{e}r", {\it Int. J. Theor. Phys.}, {\bf 39}, 479, 2000, and quant-ph/0105107.

\smallskip
\noindent
[33] {\bf Van Steirteghem, B.}, ``Quantum axiomatics: investigation of 
the
structure of the category of physical entities and Sol\'{e}r's theorem",
graduation thesis, FUND, Brussels Free University, Pleinlaan 2, 1050
Brussels.

\smallskip
\noindent
[34] {\bf Piron, C.,} ``Axiomatique Quantique", {\it Helv. Phys. Acta}, 
{\bf
37}, 439, 1964.

\smallskip
\noindent
[35] {\bf Amemiya, I. and Araki, H.}, ``A remark of Piron's paper", Publ. 
Res.
Inst. Math. Sci. {\bf A2}, 423, 1966.

\smallskip
\noindent
[36] {\bf Zierler, N.}, ``Axioms for non-relativistic quantum mechanics", 
{\it
Pac. J. Math.}, {\bf 11}, 1151, 1961.

\smallskip
\noindent
[37] {\bf Varadarajan, V.,} {\it Geometry of Quantum Theory}, Von 
Nostrand,
Princeton, New Jersey, 1968.

\smallskip
\noindent
[38] {\bf Piron, C.,} {\it Foundations of Quantum Physics}, Benjamin, 
reading,
Massachusetts, 1976.

\smallskip
\noindent
[39] {\bf Wilbur, W.}, ``On characterizing the standard quantum logics", 
{\it
Trans. Am. Math. Soc.}, {\bf 233}, 265, 1977.

\smallskip
\noindent
[40] {\bf Keller, H. A.}, ``Ein nicht-klassicher Hilbertsher Raum", {\it 
Math.
Z.}, {\bf 172}, 41, 1980.

\smallskip
\noindent
[40] {\bf Aerts, D.,} ``The one and the many", Doctoral Thesis, Brussels Free University, Pleinlaan 2, 1050 Brussels, 1981. 

\smallskip
\noindent
[41] {\bf Aerts, D.,} ``Description of many physical entities  without 
the
paradoxes encountered in quantum mechanics", {\it Found. Phys.\/}, {\bf
12}, 1131, 1982.

\smallskip
\noindent
[42] {\bf Aerts, D.,} ``Classical theories and non classical theories as 
a
special case of a more general theory", {\it J. Math. Phys.\/} {\bf
24}, 2441,  1983.

\smallskip
\noindent
[43] {\bf Valckenborgh, F.,} ``Closure structures and the theorem of 
decomposition in classical components", {\it Tatra Mountains Mathematical 
Publications}, {\bf 10}, 75, 1997.

\smallskip
\noindent
[44] {\bf Aerts, D.,} ``The description of one and many physical 
systems",
in {\it Foundations of Quantum Mechanics\/}, eds. C. Gruber, A.V.C.P.
Lausanne, 63, 1983.

\smallskip
\noindent
[45] {\bf Aerts, D.}, ``Construction of a structure which makes it
possible to describe the joint system of a classical and a quantum 
system", {\it
Rep. Math. Phys.\/}, {\bf 20}, 421, 1984.

\smallskip
\noindent
[46] {\bf Piron, C.,} {\it M\'ecanique Quantique: Bases et applications},
Presse Polytechnique et Universitaire Romandes, Lausanne, 1990.

\smallskip
\noindent
[47] {\bf Pulmannova, S.}, ``Axiomatization of quantum logics", {\it Int. 
J.
Theor. Phys.}, {\bf 35}, 2309, 1995.

\smallskip
\noindent
[48] {\bf Sol\`{e}r, M.P.}, ``Characterization of Hilbert spaces with
orthomodular spaces", {\it Comm. Algebra}, {\bf 23}, 219, 1995.

\smallskip
\noindent
[49] {\bf Holland Jr, S.S.}, ``Orthomodularity in infinite dimensions: a
theorem of M. Sol\`{e}r", {\it Bull. Aner. Math. Soc.}, {\bf 32}, 205, 
1995.

\smallskip
\noindent
[50] {\bf Aerts, D.}, ``Construction of the tensor product for lattices 
of
properties of physical entities", {\it J. Math. Phys.\/}, {\bf 25},
1434, 1984.

\smallskip
\noindent
[51] {\bf Aerts, D. and Daubechies, I.}, ``Physical justification for
using the tensor product to describe two quantum systems as one joint
system", {\it Helv. Phys. Acta} {\bf 51} , 661, 1978.

\smallskip
\noindent
[52] {\bf Jauch, J.}, {\it Foundations of Quantum Mechanics},
Addison-Wesley, Reading, Mass, 1968.

\smallskip
\noindent
[53] {\bf Cohen-Tannoudji, C., Diu, B. and Lalo{\"e}, F.},  {\it 
M{\'e}canique
Quantique, Tome I}, Hermann, Paris, 1973.

\smallskip
\noindent
[54] {\bf Aerts, D.}, ``A mechanistic classical laboratory situation
violating the Bell inequalities with
$\sqrt{2}$, exactly `in the same way' as its violations by the EPR
experiments", {\it Helv. Phys. Acta,} {\bf 64}, 1, 1991.

\end{document}